\documentclass[11pt]{article} 

\pdfoutput=1

\usepackage{times}
\usepackage{amsmath,amsthm,amsfonts,amssymb,calc,mathtools}
\usepackage{color}
\usepackage{algorithm}
\usepackage[noend]{algorithmic}
\usepackage{graphicx}
\usepackage{psfrag}
\usepackage{natbib}
\usepackage{epstopdf}
\usepackage{fullpage}
\usepackage{hyperref}
\usepackage{authblk}
\usepackage[caption=false]{subfig}
\usepackage{xcolor}
\usepackage{diagbox}
\usepackage{relsize}
\usepackage{tikz}
\usepackage{verbatim}
\usepackage{soul}
\usepackage{cancel}
\usepackage[labelfont=bf]{caption}
\usetikzlibrary{shapes,arrows}
\usepackage[customcolors]{hf-tikz}

\definecolor{naranjal}{rgb}{1.0,0.85,0.7}
\sethlcolor{naranjal}

\hfsetfillcolor{naranjal}
\hfsetbordercolor{naranjal}

\hypersetup{
    pdftitle={MIXANDMIX: numerical techniques for the computation of empirical spectral distributions of population mixtures},    
    pdfauthor={Lucilio Cordero-Grande},     
    pdfkeywords={large dimensional statistics, random matrix theory, generalized Marcenko-Pastur equations, asymptotic eigenvalue distribution, numerical solutions}, 
}

\tikzstyle{block} = [draw, fill=cyan!20, rectangle, rounded corners=0.5cm,
    minimum height=3em, minimum width=5em]
\tikzstyle{input} = [coordinate]

\pdfpageattr {/Group << /S /Transparency /I true /CS /DeviceRGB>>} 

\tikzstyle{nosep}=[inner sep=0pt, outer sep=0pt]
\title{MIXANDMIX: numerical techniques for the computation of empirical spectral distributions of population mixtures}

\date{}
\author{L Cordero-Grande}
\affil[]{Centre for the Developing Brain and Biomedical Engineering Department\\School of Biomedical Engineering and Imaging Sciences\\King's College London, King's Health Partners, St Thomas' Hospital, London, SE1 7EH, UK}
\affil[]{\small{lucilio.cordero@kcl.ac.uk}}

\newcommand\argmin{\ensuremath{\operatornamewithlimits{argmin}}}

\newcommand{\op}{
  \mathop{
    \vphantom{\bigoplus} 
    \mathchoice
      {\vcenter{\hbox{\resizebox{\widthof{$\displaystyle\bigoplus$}}{!}{$\boxplus$}}}}
      {\vcenter{\hbox{\resizebox{\widthof{$\bigoplus$}}{!}{$\boxplus$}}}}
      {\vcenter{\hbox{\resizebox{\widthof{$\scriptstyle\oplus$}}{!}{$\boxplus$}}}}
      {\vcenter{\hbox{\resizebox{\widthof{$\scriptscriptstyle\oplus$}}{!}{$\boxplus$}}}}
  }\displaylimits 
}

\DeclareMathOperator{\tr}{tr} 

\begin{document}

\maketitle

\begin{abstract}
The MIXANDMIX (mixtures by Anderson mixing) tool for the computation of the empirical spectral distribution of random matrices generated by mixtures of populations is described. Within the population mixture model the mapping between the population distributions and the limiting spectral distribution can be obtained by solving a set of systems of non-linear equations, for which an efficient implementation is provided. The contributions include a method for accelerated fixed point convergence, a homotopy continuation strategy to prevent convergence to non-admissible solutions, a blind non-uniform grid construction for effective distribution support detection and approximation, and a parallel computing architecture. Comparisons are performed with available packages for the single population case and with results obtained by simulation for the more general model implemented here. Results show competitive performance and improved flexibility.
\end{abstract}


\section{Introduction}

\label{sec:INTR}

Random matrix theory is at the core of modern high dimensional statistical inference~\citep{Yao15} with applications in physics, biology, economics, communications, computer science or imaging~\citep{Couillet13,Paul14,Bun17}. In the large dimensional scenario, classical asymptotics where the available number of samples of a given population $N$ is much larger than the dimension $M$ are no longer valid. A key contribution in this setting is the Mar\v{c}enko-Pastur theorem~\citep{Marcenko67,Silverstein95a}, which characterizes the limiting behaviour of the empirical spectral distribution (ESD) for matrices with random entries when $N\to\infty$. Namely, there exists a fixed point equation relating the eigenvalues of the empirical and population distributions, which can be used for inference under the mediation of appropriate numerical techniques. Using this characterization, subsequent asymptotics-based inference can be performed, for instance, for dimensionality reduction, hypothesis testing, signal retrieval, classification or covariance estimation~\citep{Yao15}. In addition,~\cite{Silverstein95b} develops on ideas outlined in~\cite{Marcenko67} to analyze the support of the ESD based on the monotonicity of the inverse of the function involved in the fixed point equation. Emerging from statistical models in array processing, the extension in~\cite{Wagner12} focuses on generalizing previous results to the case where the matrix rows are independent but drawn from a collection of population distributions. Here, the relation between the population covariances and the limiting behaviour of the sample eigenvalues is governed by a system of non-linear equations and, although asymptotic sample eigenvalue confinement has also been proved~\citep{Kammoun16}, a simple description of the support is no longer at hand.

Despite the variety of potential applications, not many works have practically confronted the numerical issues involved in computing the ESD from a given population distribution. The most flexible package that we have identified has been described in~\cite{Dobriban15} (SPECTRODE), where the fixed point equation in~\cite{Silverstein95a} is transformed into an ordinary differential equation (ODE) with starting point obtained by the solution of the fixed point equation on a single point within the support. In this work the author shows that the proposed method compares favourably with straightforward fixed point solvers, both in terms of accuracy and computational efficiency. In addition, accuracy limitations of Monte Carlo simulations~\citep{Jing10} are showcased and, while recognizing their independent interest, arguments are provided about the practical limitations~\citep{Rao08} or limited applicability~\citep{Olver13} of other approaches. In our experiments, this package has revealed an exquisite accuracy and efficiency in computing the ESD and determining its support. However, the applicability or efficiency of an ODE approach may be compromised for more general models such as~\cite{Wagner12}. This is mainly due to the increased computational complexity of evaluating the Jacobian of the system of fixed point equations, which no longer depends on the eigenvalues of the populations but on a combination of their covariances. Another interesting tool, conceived to solve the more general problem of recovering the population distribution from the observed eigenvalues by means of the so called QuEST function, has been described in~\cite{Ledoit17}. Solving this problem is required, for instance, for covariance estimation. The literature review in this work points to a systematic limitation of most precedent methods, as they are only capable to obtain estimates for very particular forms of the population distribution. In addition, it introduces an interesting feature not present in~\cite{Dobriban15}, the use of a non-uniform grid with increased resolution near the support edges, which allows more efficient approximations. However, once again, this technique is designed for cases where a single nonlinear equation is to be solved while generalizations to systems of equations do not seem straightforward.

This technical note describes a series of tools that have been developed to compute the ESD for the mixture of populations case in~\citep{Wagner12}. This model or certain analogous forms, has attracted interest in areas such as telecommunications~\citep{Moustakas07,Couillet11,Dupuy11}, machine learning~\citep{Benaych-Georges16,Couillet16}, medical imaging~\citep{Cordero-Grande19} or genetics~\citep{Fan19}. Our method is based on directly solving the system of nonlinear equations. However, the results in~\cite{Dobriban15} and our own analysis, have identified certain limitations in commonly reported algorithms based on fixed point iteration solvers, so a set of technical refinements are proposed in this note. These include the use of Anderson mixing to accelerate the fixed point iterations, an homotopy continuation method to prevent non-admissible solutions, a set of heuristics to detect the support of the distribution and to adapt the approximation grid to the ESD shape, and a formulation that allows for efficient computations in graphical processing units (GPU). We validate our method by comparison with~\cite{Dobriban15} and~\cite{Ledoit17}, both in terms of efficiency and accuracy. As our methods are envisaged to operate on more general models than those contemplated in~\cite{Dobriban15,Ledoit17}, they do not make use of any precomputed information about the distribution support. Nevertheless, we show that they are reliable enough and their efficiency and accuracy is comparable or superior to that in~\cite{Dobriban15,Ledoit17}. In addition, comparisons with Monte Carlo simulations show that they are also capable of providing accurate results for more general models. A \textsmaller{\textsc{MATLAB}} implementation of our approach, that we will refer to as the MIXANDMIX (Mixtures by Anderson Mixing) method, including the scripts required to replicate the experiments in this note, has been made publicly available at \url{https://github.com/mriphysics/MixAndMix/releases/tag/1.1.0}. This note is organized as follows: in~\S~\ref{sec:THEO} we review different random matrix models, in~\S~\ref{sec:METH} we describe the main features of our method, in~\S~\ref{sec:RESU} we validate the proposed technique, in~\S~\ref{sec:DISC} we discuss the implications of the obtained results and in~\S~\ref{sec:CONC} we end up with some conclusions.

\section{Theory}

\label{sec:THEO}

Consider a complex random matrix $\mathbf{X}$ of size $N\times M$. We are interested in the eigenvalue distribution of the sample covariance $\mathbf{Y}=\displaystyle N^{-1}\mathbf{X}^H\mathbf{X}$ in the asymptotic regime where both $M\to\infty$ and $N\to\infty$ but they maintain a fixed ratio $\gamma=M/N$ with $\gamma>0$. For simplicity we assume the entries of the matrix are zero mean Gaussian distributed, but keeping in mind that the literature contemplates different generalizations. Three main scenarios are contemplated:

\subsection{IID standard entries}

\label{sec:IIDS}

Within this model, that we call \emph{standard} model, we can write $\mathbf{X}^H\sim\boldsymbol{\mathcal{CN}}\displaystyle\left(\mathbf{0}_{MN},\mathbf{I}_{MN}\right)$, with $\boldsymbol{\mathcal{CN}}$ denoting a circularly symmetric complex Gaussian distribution, $\mathbf{0}_{MN}$ the $M\times N$ matrix with zero entries, and $\mathbf{I}_{MN}$ the $MN\times MN$ identity matrix.~\cite{Marcenko67} showed that in this case the distribution of eigenvalues of the sample covariance asymptotically converges to
\begin{equation}
\label{ec:MPLA}
f_{\gamma}^{\mathbf{I}}(x)=\begin{cases}\displaystyle\frac{\sqrt{(\gamma_{+}-x)(x-\gamma_{-})}}{2\pi\gamma x} & \mbox{if }\gamma_{-}\leq x\leq \gamma_{+}\\ 0 & \mbox{otherwise,}\end{cases}
\end{equation}
with $\gamma_{-}=(1-\sqrt\gamma)^2$ and $\gamma_{+}=(1+\sqrt\gamma)^2$ defining the support of the distribution. Note that if $\gamma\geq 1$, the distribution has a $1-\gamma^{-1}$ point mass ($\gamma>1$) or is locally unbounded ($\gamma=1$) at $x=0$. Thus, in this case,~\eqref{ec:MPLA} gives an explicit characterization of the ESD.

\subsection{IID rows}

\label{sec:IIDC}

In this scenario, we can write $\mathbf{X}^H\sim\boldsymbol{\mathcal{CN}}\displaystyle\left(\mathbf{0}_{MN},\boldsymbol{\Lambda}_M\otimes\mathbf{I}_N\right)$, with $\boldsymbol{\Lambda}_M$ a given population covariance matrix.~\cite{Marcenko67} and~\cite{Silverstein95a} respectively derived expressions for probabilistic and almost sure limits of the sample covariance spectrum using its Stieltjes transform
\begin{equation}
m(z)=\int_{\mathbb{R}}\frac{f_{\gamma}^{\boldsymbol{\Lambda}_M}(x)}{x-z}dx,\quad z\in\mathbb{C}\setminus\mathbb{R}
\end{equation}
for which the inversion formula would give back the ESD by
\begin{equation}
\label{ec:INST}
f_{\gamma}^{\boldsymbol{\Lambda}_M}(x)=\frac{1}{\pi}\lim_{\epsilon\to 0^{+}}\Im\left\{m(x+i\epsilon)\right\}.
\end{equation}
If, in a discrete setting, we denote the increasingly sorted eigenvalues of $\boldsymbol{\Lambda}_M$ by $\{\lambda_1,\ldots,\lambda_M\}$, when $N\to\infty$ the eigenvalue distribution of the matrix $\tilde{\mathbf{Y}}=\displaystyle N^{-1}\mathbf{X}\mathbf{X}^H$ converges to a function whose Stieltjes transform $\tilde{m}(z)$ is related to the population distribution by the fixed point equation\newline $\tilde{m}(z)=\displaystyle\left(-z+\frac{\gamma}{M}\sum_{m=1}^M\frac{\lambda_m}{1+\lambda_m \tilde{m}(z)}\right)^{-1}$ and to the Stieltjes transform of the corresponding limiting function for the spectral distribution of the sample covariance matrix $\mathbf{Y}$, $m(z)$, by $\tilde{m}(z)=\gamma m(z)+(\gamma-1)/z$. In addition,~\cite{Silverstein95b}, following the guidelines in~\cite{Marcenko67}, showed that the support of the ESD can be characterized by studying the zeros of the derivative of the inverse map, $z'(\tilde{m})$, in appropriate domains. This property, together with the analyticity of the empirical density within its support are the keys to the SPECTRODE approach in~\cite{Dobriban15}.

To clarify the connections between the different models, it is more convenient to express the previous relations in terms of auxiliary functions $e(z)=-\displaystyle\frac{1}{\gamma z\tilde{m}(z)}-\frac{1}{\gamma}$ for an equivalent fixed point equation,
\begin{equation}
\label{ec:CFST}
e(z)=\frac{1}{M}\sum_{m=1}^M\lambda_m\left(\frac{\lambda_m}{1+\gamma e(z)}-z\right)^{-1},
\end{equation}
and an expression for the Stieltjes transform of the ESD,
\begin{equation}
\label{ec:STSD}
m(z)=\frac{1}{M}\sum_{m=1}^M\left(\frac{\lambda_m}{1+\gamma e(z)}-z\right)^{-1}.
\end{equation}
We refer to this model as the \emph{single population} model.

\subsection{Independent rows}

\label{sec:INCO}

In this \emph{mixture of populations} model, the matrix is drawn from $\mathbf{X}^H\mathbin{\sim}\boldsymbol{\mathcal{CN}}\displaystyle\left(\mathbf{0}_{MN},\sum_{k=1}^{K}\boldsymbol{\Lambda}^k_M\otimes\mathbf{D}^k_N\right)$, where $\mathbf{D}^k_N$ is a diagonal indicator matrix with ones in the diagonal elements corresponding to those rows sampled according to the population covariance $\boldsymbol{\Lambda}^k_M$ and zero otherwise. The equations relating the empirical and population distributions for $N\to\infty$ have been derived in~\cite{Wagner12} also making use of the Stieltjes transform of the limiting spectral distribution $f_\gamma^{\prescript{}{K}{\boldsymbol{\Lambda}}^{}_{M}}$, $m(z)$, and the auxiliary functions $e_j(z)$, $1\leq j\leq K$. These functions are related to the population covariances by a system of nonlinear equations
\begin{equation}
\label{ec:CSST}
e_j(z)=\frac{1}{M}\tr\left(\boldsymbol{\Lambda}_j\left(\sum_{k=1}^{K}\frac{\alpha_k\boldsymbol{\Lambda}_k}{1+\gamma e_k(z)}-z\mathbf{I}_M\right)^{-1}\right),
\end{equation}
for which there is a unique solution in $\mathbb{C}\setminus\mathbb{R}^{+}$. $m(z)$ is expressed in terms of these functions as
\begin{equation}
\label{ec:SSSD}
m(z)=\frac{1}{M}\tr\left(\left(\sum_{k=1}^{K}\frac{\alpha_k\boldsymbol{\Lambda}_k}{1+\gamma e_k(z)}-z\mathbf{I}_M\right)^{-1}\right),
\end{equation}
with $\alpha_k=\tr(\mathbf{D}^k)/N$. Note that~\eqref{ec:CSST} and~\eqref{ec:SSSD} reduce to~\eqref{ec:CFST} and~\eqref{ec:STSD} when $K=1$. There are two main limitations to extend the SPECTRODE method to this setting. First, we are unaware of studies characterizing the support of the measures inducing $e_j(z)$ by means of some analogue to~\cite{Silverstein95b}. Second, both potential extensions of support characterizations or usage of ODE solvers would require the Jacobian of~\eqref{ec:CSST}, which involves additional matrix multiplications, with a penalty in computational efficiency. Thus, we have focused on developing a reliable method to solve the system~\eqref{ec:CSST} not requiring the Jacobian.

\section{Methods}

\label{sec:METH}

In this Section we describe our numerical solver for the system of equations in~\eqref{ec:CSST}.

\subsection{Support detection}

\label{sec:SUDE}

When $\gamma\to 0$ the ESD tends to the population distribution for the standard and single population models, while for the mixture of populations it is governed by an effective single population distribution $f_\gamma^{\prescript{}{K}{\boldsymbol{\Lambda}}^{}_{M}}=f_0^{\overline{\boldsymbol{\Lambda}}_M}$ with $\overline{\boldsymbol{\Lambda}}_M=\displaystyle\sum_{k=1}^K\alpha_k\boldsymbol{\Lambda}_M^k$, so that we have $f_0^{\overline{\boldsymbol{\Lambda}}_M}(x)=\displaystyle\frac{1}{M}\sum_{m=1}^M\delta(x-\overline{\lambda}_m)$, with $\overline{\lambda}_m$ the $m$-th eigenvalue of $\overline{\boldsymbol{\Lambda}}_M$. Thus, in this limiting case the ESD support only includes the eigenvalues of an equivalent single population distribution. On the other side, for $\gamma>0$, $[a,b]=\left [t^{-1}(1-\sqrt{\gamma})^2\displaystyle\min_k\lambda_1^k,t(1+\sqrt{\gamma})^2\max_k\lambda_M^k\right ]$, with $\lambda_m^k$ denoting the $m$-th eigenvalue of $\boldsymbol{\Lambda}_M^k$ and $t>1$, provide lower and upper bounds on the limiting support.

To consider these two extreme cases, we group and sort the set of $P=M(K+1)$ eigenvalues $\lambda_p=\{\lambda_m^k,\overline{\lambda_m}\}$, $1<k<K$, $1<m<M$. Our method attends to the overlap of the set of intervals
\begin{equation}
\label{ec:OVER}
[a_p,b_p]=[t^{-1}(1-\sqrt{\gamma})^2\lambda_p,t(1+\sqrt{\gamma})^2\lambda_p],
\end{equation}
i.e., the intervals generated by the maximum possible spectral dispersion of each grouped eigenvalue, including a numerical safety margin $t$, which we have set to $t=1.001$ (and constraining $|\gamma-1|\geq 10^{-9}$ to avoid the potential singularity at $x=0$ when $\gamma=1$).

The support of the limiting distribution will be contained in $[a,b]$, as this interval considers the maximum possible eigenvalue dispersion as produced by minorizing ($a$) and majorizing ($b$) point mass population distributions governed by~\eqref{ec:MPLA}. On the contrary, there is no guarantee for $\displaystyle\bigcup_p [a_p,b_p]$ to contain the whole support of the limiting distribution. However, the intervals in~\eqref{ec:OVER} serve to cover the $\gamma\to 0$ regime, which originates challenging highly localized support intervals, as in this case our construction guarantees that the support will be contained in the defined intervals. Although we have empirically observed cases where part of the support lies on $[a,b]\setminus\displaystyle\bigcup_p [a_p,b_p]$, we have never observed that these intervals contain any of the support edges. In this situation, the adaptive regridding procedure in~\S~\ref{sec:ADRE} should generally be enough to interpolate the distribution estimates on $[a,b]\setminus\displaystyle\bigcup_p [a_p,b_p]$.

Lack of overlap among adjacent intervals is detected by checking the condition $b_p<a_{p+1}$. Assuming this is observed $I\leq P-1$ times, we can get a partition into $I+1$ segments, each one induced by $P_i$ eigenvalues. Each of these support segments is gridded using a total of $\max(M^{\text{(o)}}P_i,M^{\text{(i)}})$ points, with $M^{\text{(i)}}$ the minimum number of points per segment and $M^{\text{(o)}}\leq M^{\text{(i)}}$ the minimum ratio of points per number of grouped eigenvalues. Importantly, to improve detectability, by noting the multiplicative dependence of the spectral dispersion width with the spectral location in~\eqref{ec:OVER}, gridding is performed uniformly in logarithmic units. Then, we can call the solver of~\eqref{ec:CSST} (to be described in~\S~\ref{sec:HOMO}), and compute~\eqref{ec:SSSD} and~\eqref{ec:INST} over the defined grid locations. 

The output of this first step is a set of estimates for the distribution in a non-uniform grid $\mathbf{x}=x_p$ with $1\leq p\leq P_0$ and $M^{\text{(o)}}P\leq P_0\leq M^{\text{(i)}}P$. The lower and upper bounds on the number of grid points, $M^{\text{(o)}}P$ and $M^{\text{(i)}}P$, are respectively attained for large and small enough $\gamma$. Fixing $M^{\text{(o)}}=3$ and $M^{\text{(i)}}=15$ has detected at least a single point within all the segments conforming the support of the distribution for the range of problems studied in~\S~\ref{sec:RESU}. For problems where the total geometric multiplicity of the discretized population covariances is given by $M^{\text{(u)}}$, equivalent expressions of the problem can be obtained for any $M=SM^{\text{(u)}}$ with $S\in\mathbb{N}_{>0}$. In our experiments $S$ has been selected by defining a minimum number of grid points to approximate the densities by $M^{\text{(m)}}$, and making $M=\max(M^{\text{(u)}},M^{\text{(m)}})$ with $M^{\text{(m)}}=100$.

\subsection{Adaptive regridding}

\label{sec:ADRE}

Additional points are added to the grid by pursuing $g'(x)\sqrt{xf''(x)}=c$, with $g(x)$ the grid mapping function and $c$ a constant. This non-uniform grid construction criterion is based on both the second order derivative of the density $f''(x)$ and the grid value $x$. The first feature, $f''(x)$, favours the allocation of grid points near the support edges, in accordance to the $\sqrt{x-x_0}$ behaviour of the distribution at the boundaries~\citep{Silverstein95b}, as well as in those areas where linear interpolation results in larger approximation errors. This is similar in spirit to the arcsine criterion in~\cite{Ledoit17} but does not use any prior information about the support edges as it is not available in the mixture of populations model. The second feature, $x$, favours the allocation of grid points near the upper edge of the support, which could be important for applications related to signal detection~\citep{Nadakuditi14,Dobriban19}. After $P_l=R_lP_0$ points are added to the grid, the solver for $f(\mathbf{x})$ is called on the new set of points to allow for an update of the $f''(x)$ values to be used at the next iterative regridding step. This whole process is repeated $L$ times, so $R_l$, $1\leq l\leq L$ control the final resolution of the distribution computations. The parameters by default in our implementation are $R_l=1$ $\forall l$ and $L=1$.

\subsection{Homotopy continuation}

\label{sec:HOMO}

The calculation of the ESD involves a pass to the limit in~\eqref{ec:INST} as the Stieltjes transform does not converge in the real line. In addition, the solution of~\eqref{ec:CSST} is not unique on the real line. Numerically, this may provoke spurious fixed point convergence when the current solution is far away from the optimum and the computations are being performed in locations that are close to the real line. To prevent these situations, we have emulated~\eqref{ec:INST} by homotopy continuation. We start by obtaining an approximate solution for~\eqref{ec:CSST} in a grid given by $\mathbf{z}=\mathbf{x}+\boldsymbol{\xi}^2i$, with $\boldsymbol{\xi}=\xi^0\mathbf{1}_{P_l}$ for big enough $\xi^0$ common for all $1\leq p\leq P_l$. At each iteration $i$ we compute $\varepsilon^i_p=\displaystyle\max_k|e^i_k(x_p)-e^{i-1}_k(x_p)|$, where the updates on $e$ are to be described in~\S~\ref{sec:ANDE}. Considering that, as discussed in~\cite{Dobriban15}, to obtain an accuracy of at least $\epsilon$ for the distribution $f(x)$, we need to solve the system of equations in a complex grid given by $x+i\epsilon^2$, we can perform the update $\xi^{j+1}_p=\max(\xi^j_p/\beta,\epsilon)$ whenever $\varepsilon^i_p\leq\varepsilon^{i-1}_p$. The iterations at the grid location indexed by $p$ are terminated when the prescribed accuracy is reached, namely when $\xi^j_p=\epsilon$ and $\varepsilon^i_p<\epsilon$. In our experiments we have used $\xi^0=1$ and $\beta=10$, for which we have observed robust and efficient performance. 

\subsection{Anderson acceleration}

\label{sec:ANDE}

The experiments in~\cite{Dobriban15} showed that when solving the IID rows problem in~\S~\ref{sec:IIDC} by a straightforward fixed point algorithm, in our context when performing the updates on $e(z)$ directly using~\eqref{ec:CFST}, convergence is often very slow, so they reported situations where their SPECTRODE code could be $1000\times$ quicker while simultaneously obtaining $1000\times$ higher accuracy. In this note we show that this apparent limitation of the fixed point iterates can be overcome by using techniques to accelerate their convergence. As discussed in~\S~\ref{sec:INCO}, the accelerated convergence achievable by methods requiring the Jacobian of the fixed point identities may not compensate for the increased cost per iteration involved in computing the Jacobian. Thus, we have resorted to Anderson mixing~\citep{Anderson65}, a technique not requiring explicit Jacobian calculations that has demonstrated good practical performance, in occasions providing competitive results when compared to gradient-based approaches~\citep{Ramiere15}.

Considering a given multidimensional fixed point mapping $\mathbf{g}(\mathbf{e})$ such as~\eqref{ec:CSST}, Anderson iterations are computed as
\begin{equation}
\mathbf{e}^{i+1}=\mathbf{g}(\mathbf{e}^{i})-\sum_{q=1}^{Q_i}(\mathbf{g}(\mathbf{e}^{i-Q_i+q})-\mathbf{g}(\mathbf{e}^{i-Q_i+q-1}))\nu^i_q,
\end{equation}
with $Q_i$ denoting the number of iterations whose history is used for the update at iteration $i$ and $\boldsymbol{\nu}^i=(\nu^i_1,\ldots,\nu^i_{Q_i})^T$ obtained by solving a linear least squares problem involving the fixed point updates $\mathbf{h}^i(\mathbf{e}^i)=\mathbf{g}(\mathbf{e}^i)-\mathbf{e}^i$ and their differences $\Delta\boldsymbol{h}^i=\boldsymbol{h}^{i}-\boldsymbol{h}^{i-1}$ arranged in a $K\times Q_i$ matrix $\Delta\boldsymbol{H}^i=[\Delta\boldsymbol{h}^{i-Q_i+1},\ldots,\Delta\boldsymbol{h}^i]$. Due to the potential ill-posedness of this system, we have actually solved a damped version~\citep{Scieur19,Henderson19}
\begin{equation}
\boldsymbol{\nu}^i=\argmin_{\boldsymbol{\nu}}\|\boldsymbol{h}^i-\Delta\boldsymbol{H}^i\boldsymbol{\nu}\|_2^2+\lambda^i\|\boldsymbol{\nu}\|_2^2,
\end{equation}
with damping parameter given by $\lambda^i=\displaystyle 0.1\max_{k,q}|\Delta H^i_{k,q}|$. We have set $Q_i=\min(2,i-1)$ on the basis of our empirical testing and in agreement with the experimental results in~\cite{Ramiere15}.

\subsection{GPU acceleration}

\label{sec:GPUA}

GPU-based implementations of the SPECTRODE method~\citep{Dobriban15} appear involved due to the sequential nature of ODE solvers. In contrast, acceleration of the ESD computation in the QuEST method~\citep{Ledoit17} seems more plausible as it solves for a zero of a function independently for the different grid locations, but the authors have not discussed this aspect. Our code has been architectured so that most demanding routines support both CPU and GPU based parallel computations. This includes the parallel computation of the solutions of the system of equations in~\eqref{ec:CSST} for the different grid locations but also the parallel computation of different ESDs, required, for instance, in patch-based image denoising applications~\citep{Cordero-Grande19}.

\section{Results}

\label{sec:RESU}

In this Section we first validate the beneficial effects of Anderson acceleration and homotopy continuation in ESD computations (\S~\ref{sec:BEFU}), then compare our method to the SPECTRODE and QuEST proposals in those regimes in which they can operate (\S~\ref{sec:COLI}) and finally provide some results on the application of our technique to the mixture of populations model (\S~\ref{sec:MIPO}). Unless otherwise stated experiments are performed using CPU computations.

\subsection{Validation of introduced refinements}

\label{sec:BEFU}

\begin{figure}[!htb]
\begin{minipage}{.32\textwidth}
\includegraphics[width=\textwidth]{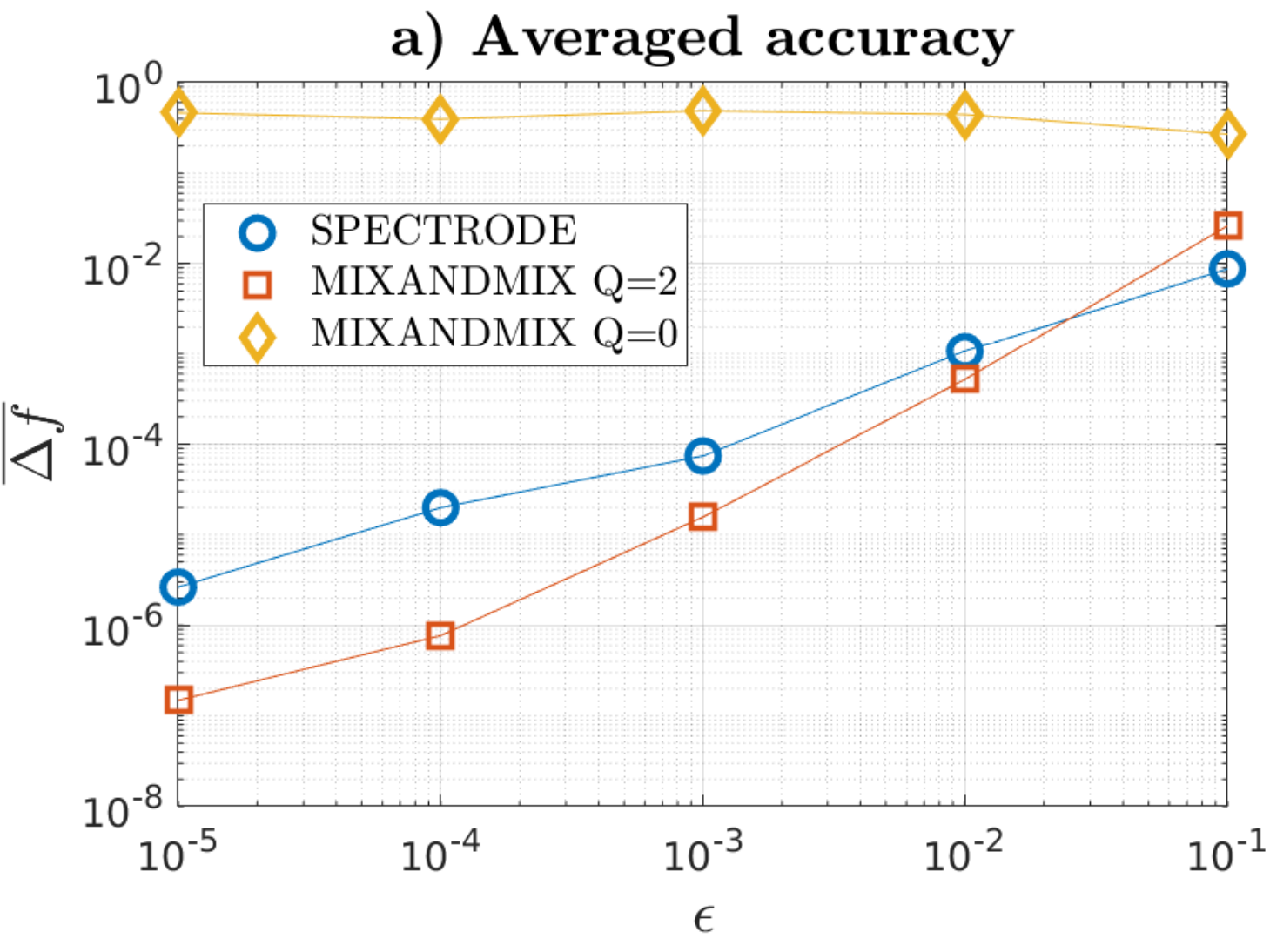}
\end{minipage}
\begin{minipage}{.32\textwidth}
\includegraphics[width=\textwidth]{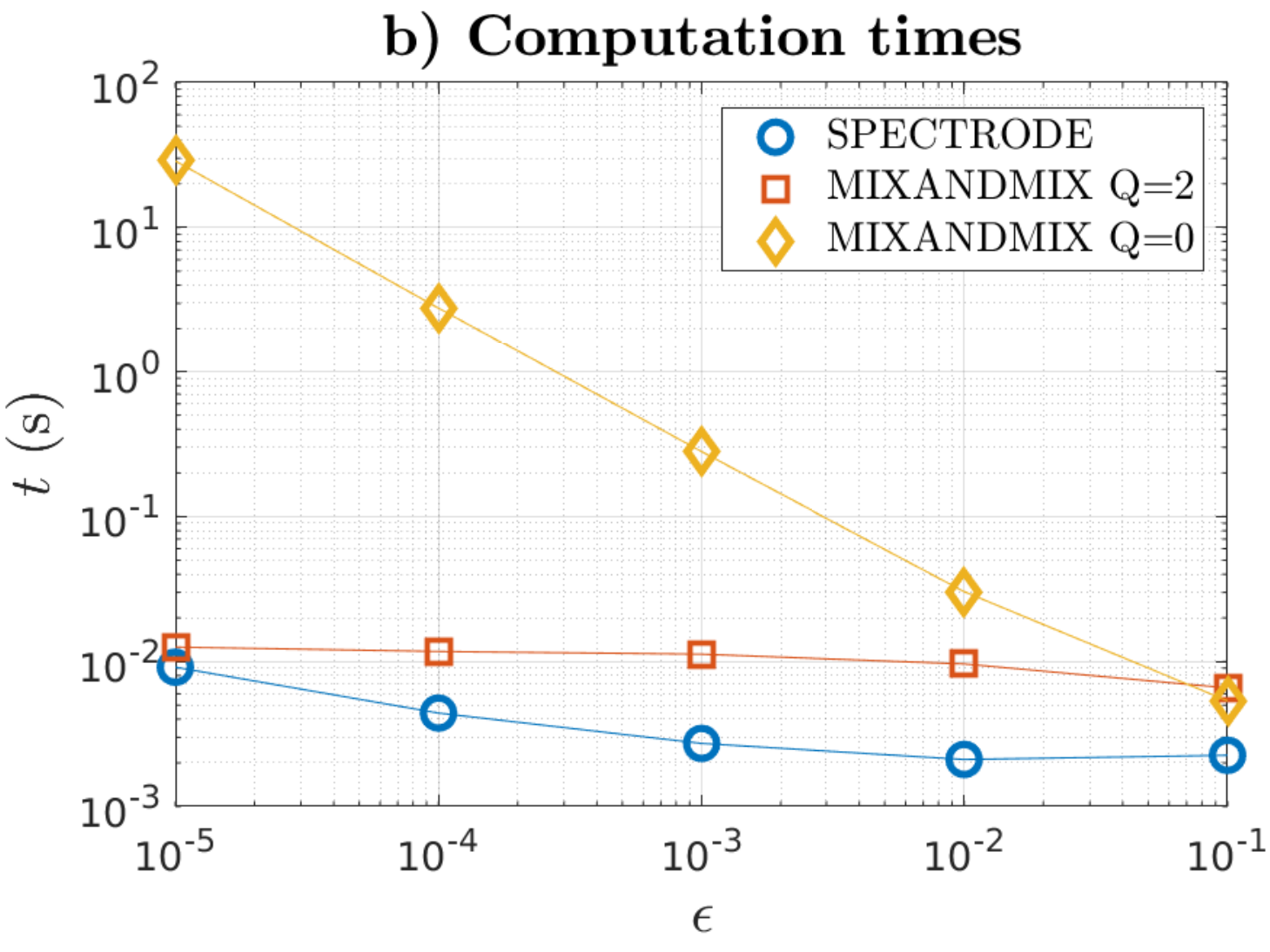}
\end{minipage}
\begin{minipage}{.32\textwidth}
\includegraphics[width=\textwidth]{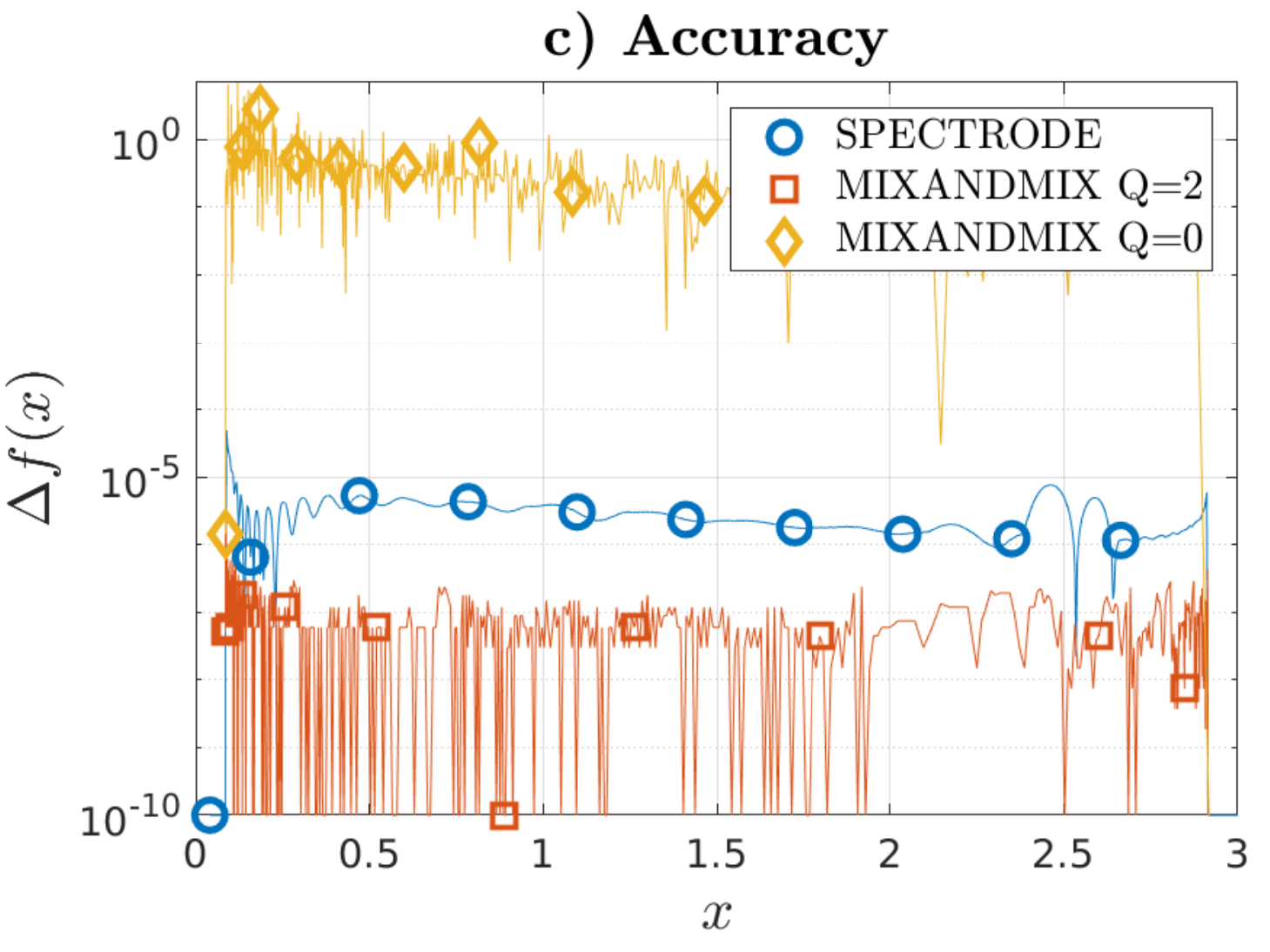}
\end{minipage}
\caption{\textbf{a)} Averaged accuracy of different methods $\overline{\Delta f}=\displaystyle\sum_{p=1}^{P}|\hat{f}_{0.5}^{\mathbf{I}}(x_p)-f_{0.5}^{\mathbf{I}}(x_p)|/P$ with $\hat{f}$ denoting the computed density. \textbf{b)} Computation times $t$. \textbf{c)} Accuracy $\Delta f(x)=|\hat{f}_{0.5}^{\mathbf{I}}(x)-f_{0.5}^{\mathbf{I}}(x)|$ throughout the support (case $\epsilon=10^{-5}$).}
\label{fig:FIG1}
\end{figure}

In~\cite{Dobriban15} an experiment was performed illustrating the limitations of fixed point iterations to obtain accurate results in ESD calculations. Their method is compared with a fixed point iteration solver for the standard model in~\S~\ref{sec:IIDS}, using the closed form density in~\eqref{ec:MPLA} with $\gamma=0.5$ to assess the accuracy. Here we repeat that experiment adding our Anderson acceleration technique to the fixed point solver. The results are presented in Fig.~\ref{fig:FIG1}. First, we have been able to replicate the results in~\cite{Dobriban15}; the fixed point algorithm, grossly equivalent to the MIXANDMIX implementation with $Q=0$, i.e., without Anderson mixing, is only able to provide very moderate accuracies, despite being run for $1/\epsilon$ iterations. However, when introducing the Anderson acceleration scheme, the results are dramatically better, with MIXANDMIX and SPECTRODE demonstrating comparable performance. In this experiment the curves show a slightly better accuracy (Fig.~\ref{fig:FIG1}a) and worse computational efficiency (Fig.~\ref{fig:FIG1}b) for MIXANDMIX, but this should be taken with caution as these tests have been conducted without considering the influence of grid sizes on the approximation, which will be taken into account in the experiments in~\S~\ref{sec:COLI}. In addition, we show (Fig.~\ref{fig:FIG1}c) that despite the accuracy obtained by a straightforward fixed point algorithm is poor everywhere within the support, the accuracy curve after Anderson mixing remains below the SPECTRODE curve almost everywhere.

\begin{figure}[!htb]
\begin{minipage}{.33\textwidth}
\includegraphics[width=\textwidth]{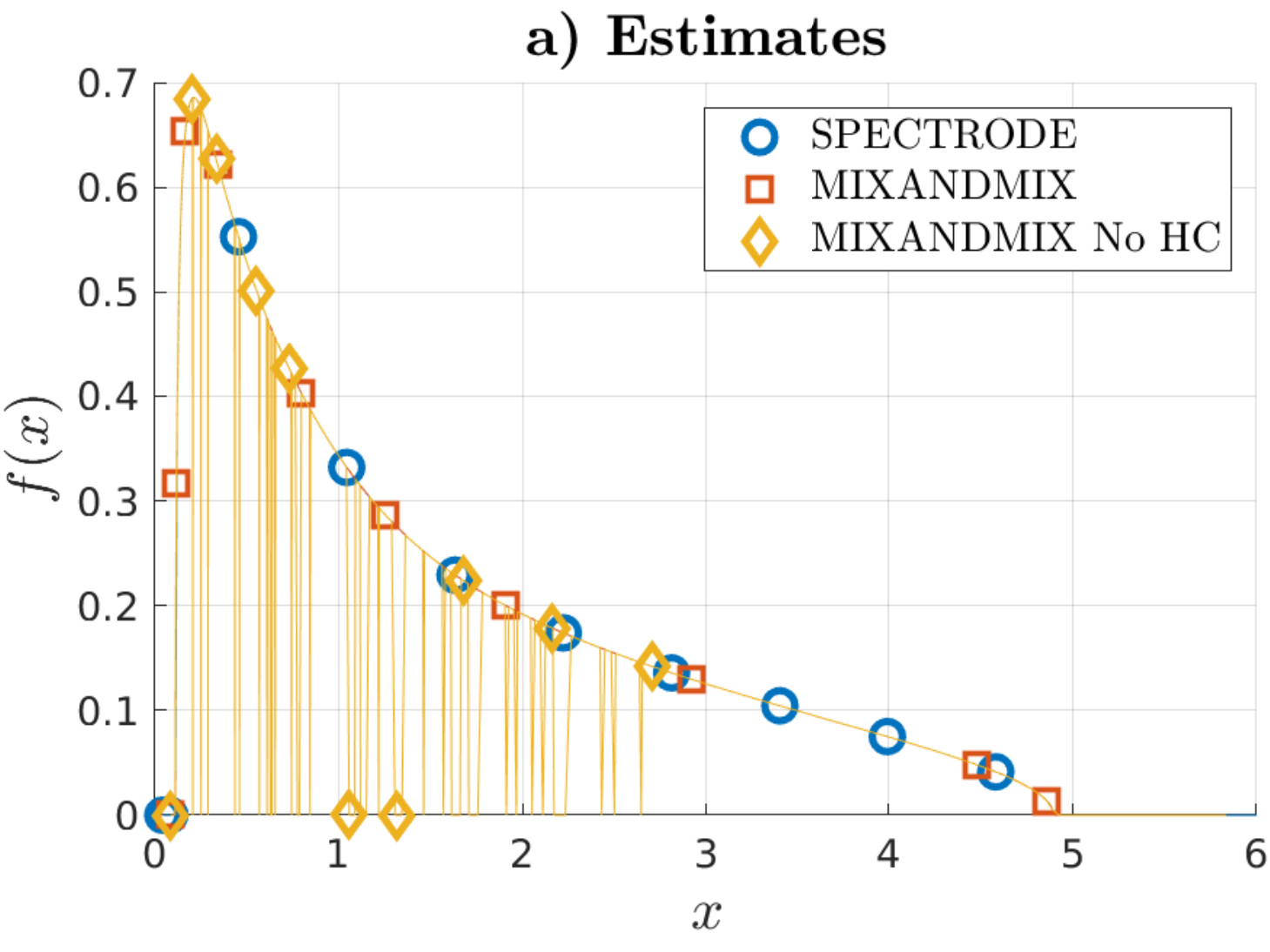}
\end{minipage}
\begin{minipage}{.66\textwidth}
\includegraphics[width=0.49\textwidth]{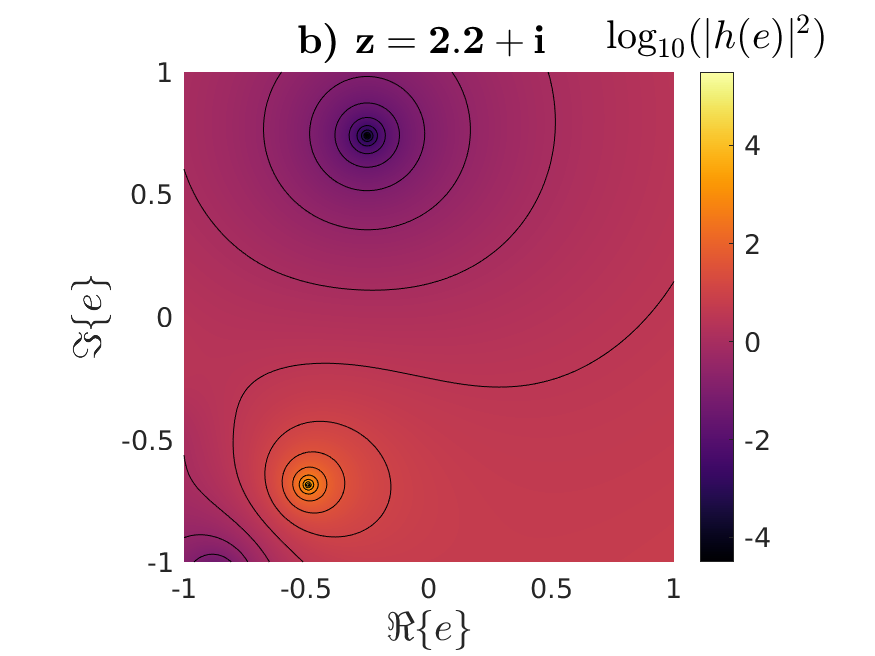}
\includegraphics[width=0.49\textwidth]{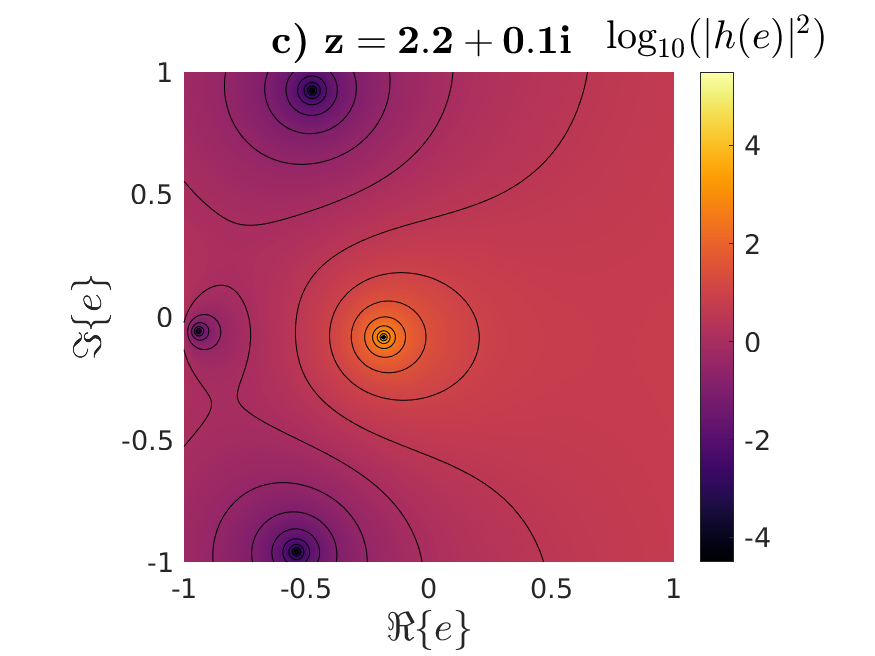}\\
\includegraphics[width=0.49\textwidth]{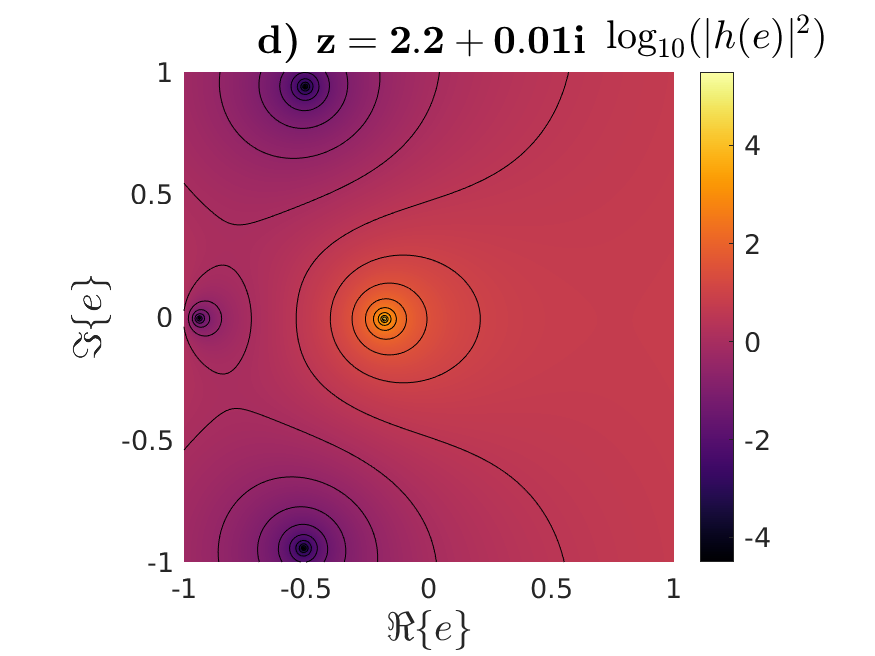}
\includegraphics[width=0.49\textwidth]{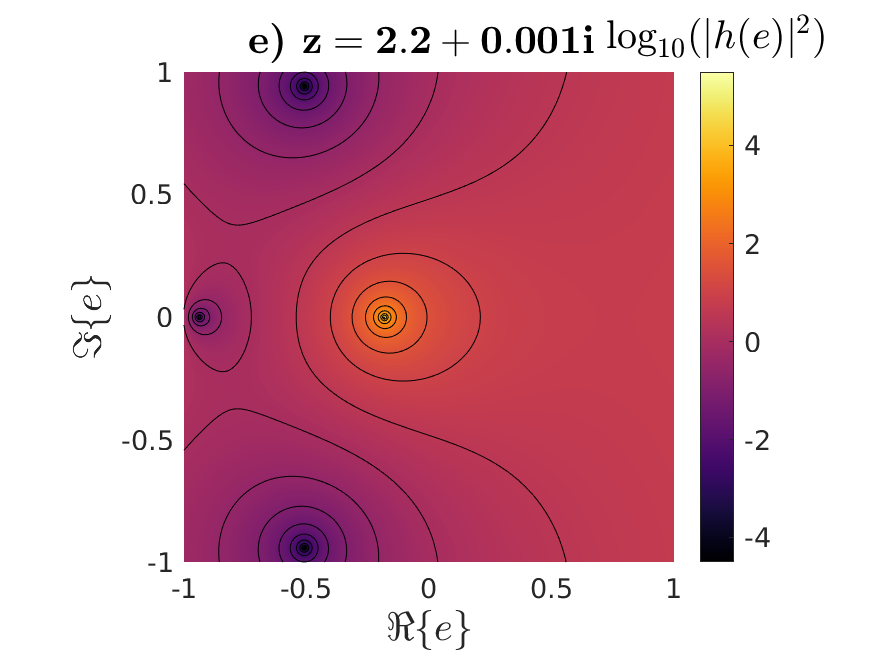}
\end{minipage}
\caption{\textbf{a)} Calculated ESDs when using the SPECTRODE method, MIXANDMIX including homotopy continuation and MIXANDMIX without homotopy continuation. \textbf{b-e)} Base $10$ logarithm of the squared magnitude of the fixed point update, i.e., $\log_{10}(|h(e)|^2)$, together with corresponding isolines at \textbf{b)} $z=2.2+1i$, \textbf{c)} $z=2.2+0.1i$, \textbf{d)} $z=2.2+0.01i$ and \textbf{e)} $z=2.2+0.001i$.}
\label{fig:FIG2}
\end{figure}

In Fig~\ref{fig:FIG2}a we compare the results of our method without and with homotopy continuation to those of SPECTRODE. The SPECTRODE method and ours with homotopy continuation are observed to overlap at the scale of the plot. However, when running MIXANDMIX without homotopy continuation, we can see that there exist some grid points for which the computations spuriously converge to the zero solution. We know this solution is infeasible because it provokes discontinuities in the distribution, which contradicts its expected analyticity properties~\citep{Silverstein95b}. To illustrate the reasons for these numerical issues, we have taken a grid point corresponding to one of these infeasible results, $x=2.2$. For this point, Figs.~\ref{fig:FIG2}b-e show the squared magnitude of the residuals of the fixed point maps, $|h(e)|^2$, at different complex plane values of the auxiliary function $e(z)=e(x+\delta i)$ as we approach the real line with $\delta=\{1,0.1,0.01,0.001\}$. First, for $z=x+1i$ there is a unique minimum in $\mathbb{C}^{+}$ whose basin of attraction covers the whole of $\mathbb{C}^{+}$. As we decrease $\delta$ (see for instance $z=x+0.1i$) we can track this minimum in a neighborhood of its previous location and check that it is still the only one in the upper half of the complex plane. However, a new local minimum has emerged in the lower half of the plane but so close to the real line that its basin of attraction extends to the upper half. As we keep decreasing $\delta$, the basin of attraction of this minimum in $\mathbb{C}^{+}$ gets bigger; however, by analyticity there has to be an area around the global optimum for which the method should still converge to the global optimum, which can be ensured by homotopy continuation. This explains the problem we are observing in the left hand side: the fixed point algorithm has entered the basin of attraction of the minimum in the lower half and it has not been able to escape from this area. In addition, the location of the attractor explains why the spurious distribution value obtained by the fixed point algorithm is generally pushed to $0$ when failing to converge to the right optimum.

\subsection{Comparison with the literature}

\label{sec:COLI}

\begin{figure}[!htb]
\begin{minipage}{.33\textwidth}
\includegraphics[width=\textwidth]{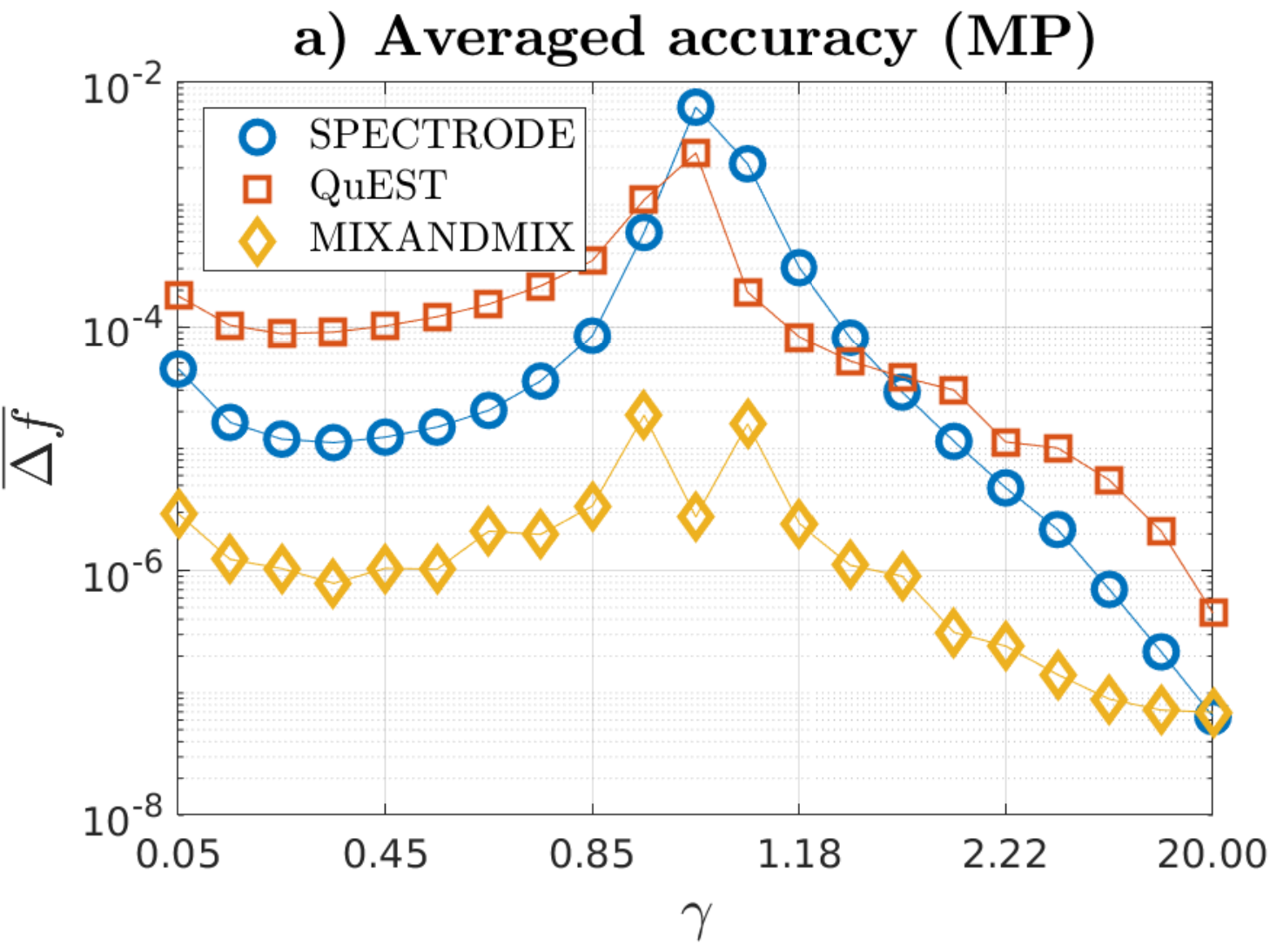}
\end{minipage}
\begin{minipage}{.33\textwidth}
\includegraphics[width=\textwidth]{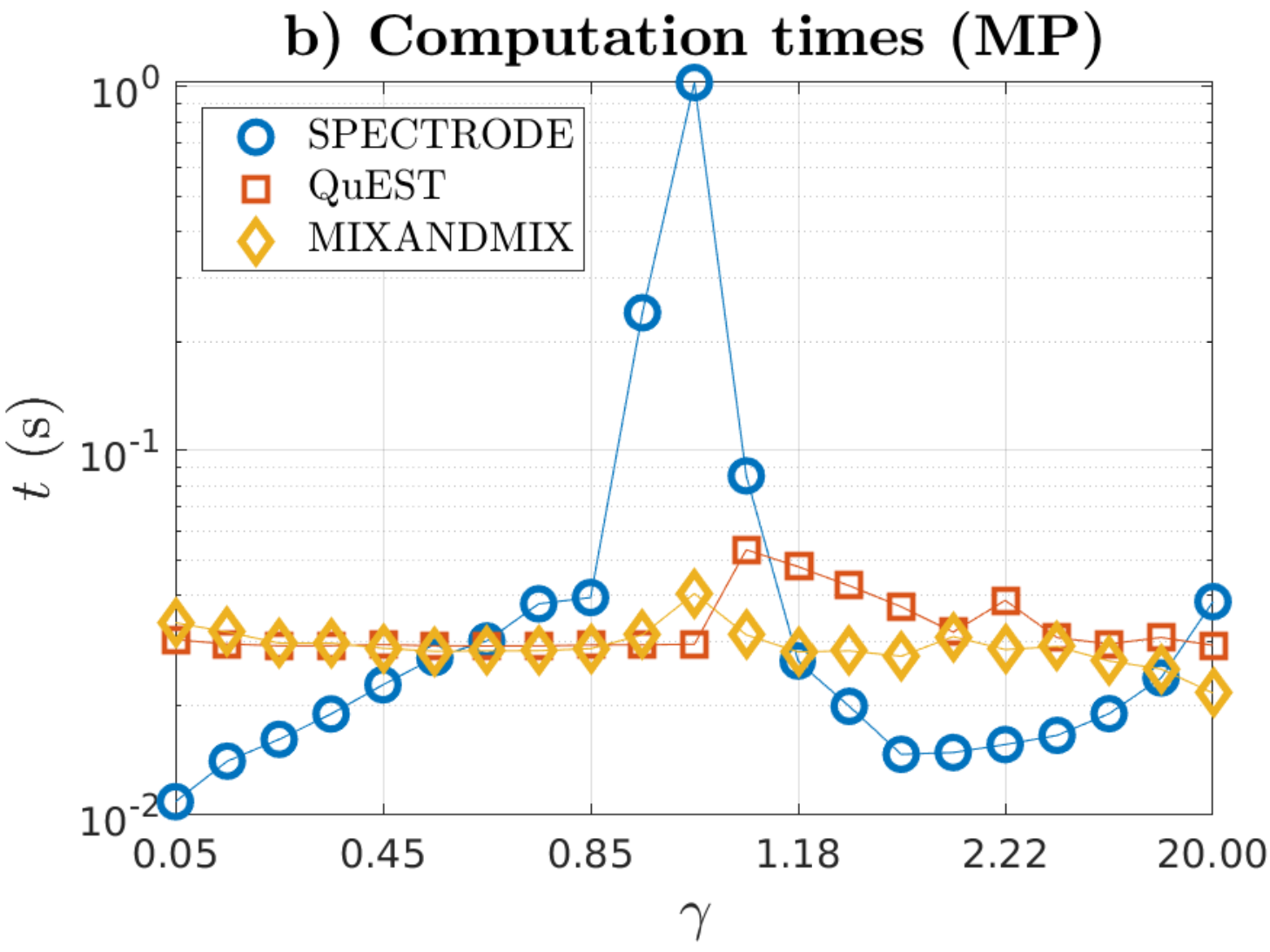}
\end{minipage}
\begin{minipage}{.33\textwidth}
\includegraphics[width=\textwidth]{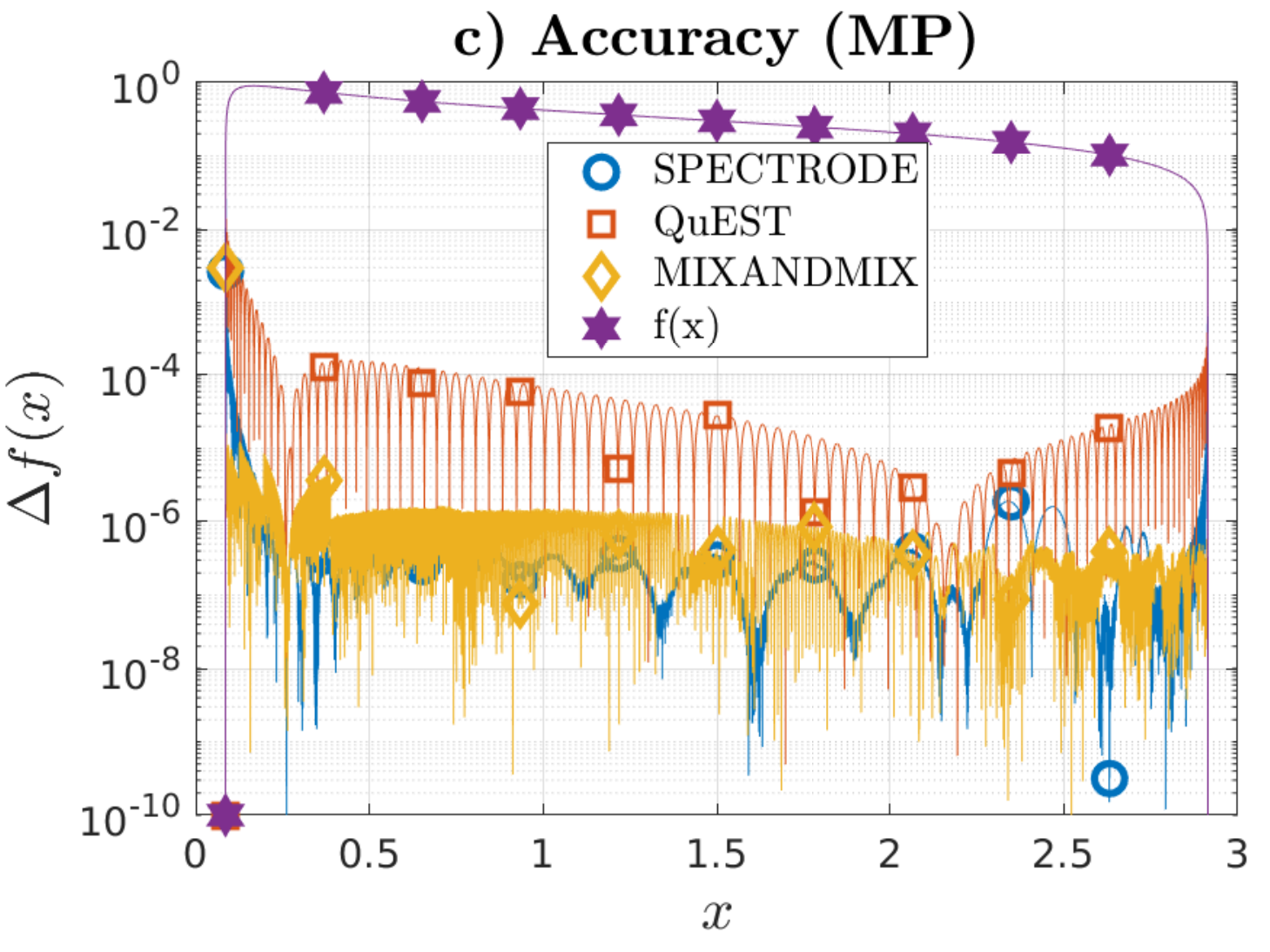}
\end{minipage}
\begin{minipage}{.33\textwidth}
\includegraphics[width=\textwidth]{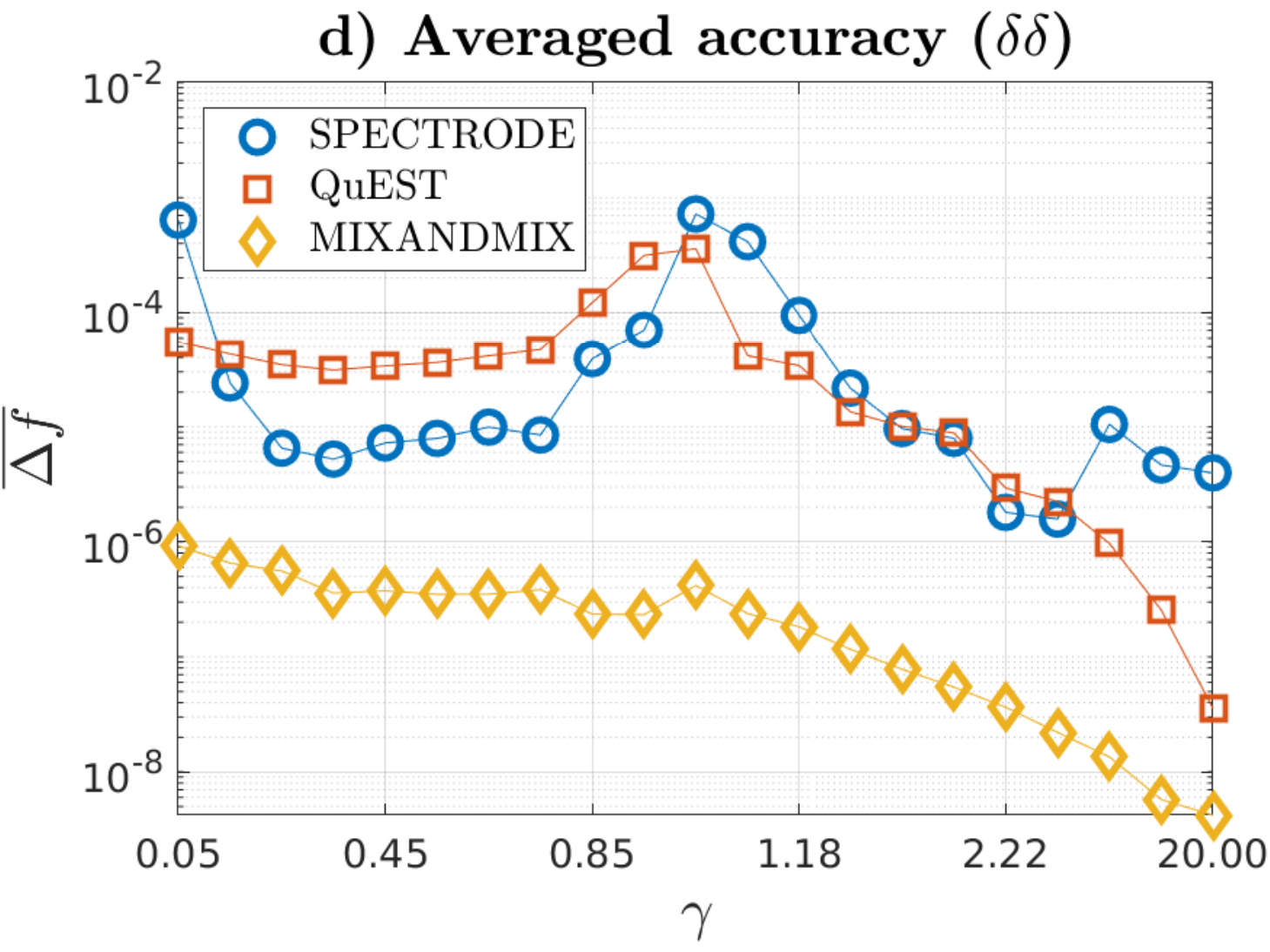}
\end{minipage}
\begin{minipage}{.33\textwidth}
\includegraphics[width=\textwidth]{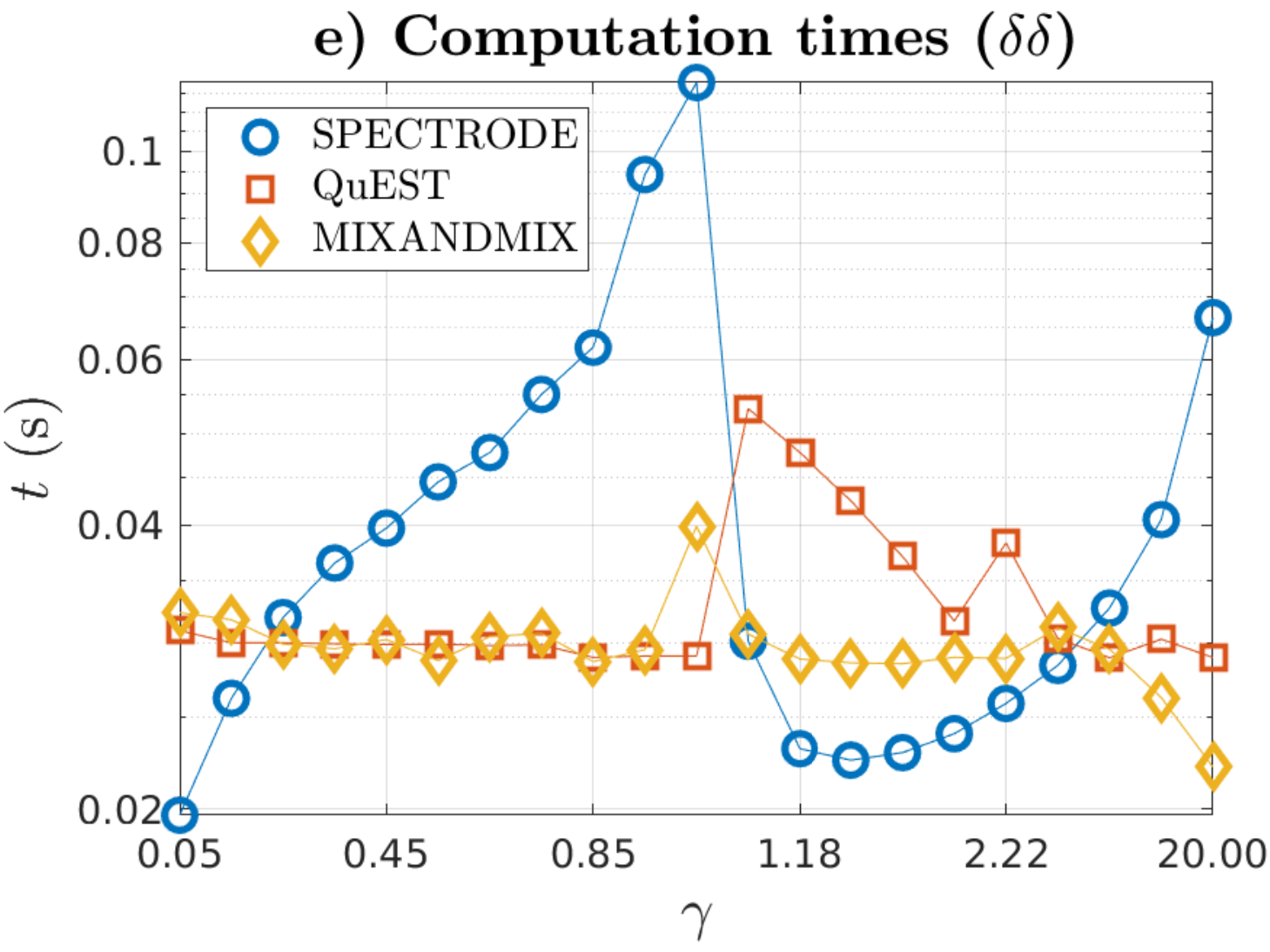}
\end{minipage}
\begin{minipage}{.33\textwidth}
\includegraphics[width=\textwidth]{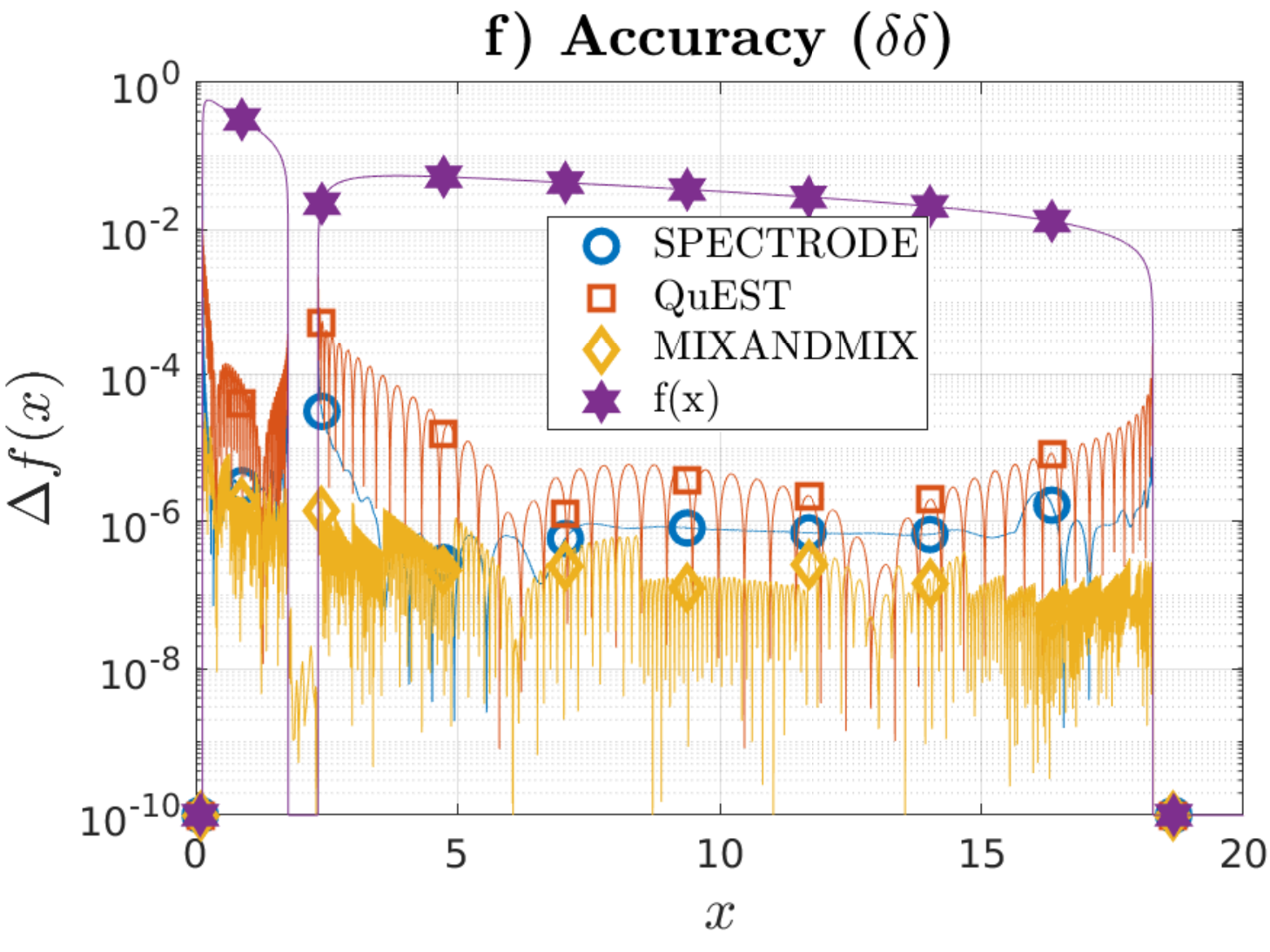}
\end{minipage}
\caption{\textbf{a,d)} Averaged accuracy, \textbf{b,e)} computation times, and \textbf{c,f)} accuracy throughout the domain ($\gamma=0.5$) for \textbf{a-c)} the MP problem and \textbf{d-f)} the $\delta\delta$ problem with equiprobable $\boldsymbol{\lambda}=\{1,8\}$.}
\label{fig:FIG3}
\end{figure}

In Fig.~\ref{fig:FIG3} we compare the accuracy and computational efficiency of MIXANDMIX with the SPECTRODE and QuEST approaches for two population distributions that admit a closed form expression for the ESD. In Figs.~\ref{fig:FIG3}a,b, we show respectively the averaged accuracy and computation times for a set of aspect ratios ranging from $\gamma=0.05$ to $\gamma=0.95$ in steps of $0.1$, $\gamma=1$, and reciprocal $1/\gamma$ ranging from $\gamma=0.05$ to $\gamma=0.95$ in steps of $0.1$ for the standard distribution (MP) in~\eqref{ec:MPLA}. Corresponding accuracies throughout the support together with the gold standard density (in a logarithmic scale) are shown in Fig.~\ref{fig:FIG3}c for the $\gamma=0.5$ case. Analogous plots are provided in Figs.~\ref{fig:FIG3}d-f for a two-delta ($\delta\delta$) distribution with equiprobable eigenvalues at $\lambda_1=1$ and $\lambda_2=8$, for which the ESD can be obtained by solving a third order polynomial equation~\citep{Dobriban15,Rao08}. The SPECTRODE and MIXANDMIX approaches have been run with an accuracy parameter providing similar computation times than those of the QuEST method using $100$ grid points, which corresponds to $\epsilon=10^{-6}/\epsilon=10^{-5}$ (MP, $\gamma<1/\gamma\geq 1$) and $\epsilon=10^{-5}/\epsilon=10^{-4}$ ($\delta\delta$, $\gamma<1/\gamma\geq 1$) for SPECTRODE and $\epsilon=10^{-5}$, $L=3$ (both) for MIXANDMIX. To account for the relative grid complexities of different methods, linearly interpolated densities are compared with close-form solutions in a uniform grid comprised of $10000$ evenly distributed points along the support. MIXANDMIX is roughly $2$ and $1$ orders of magnitude more accurate than QuEST and SPECTRODE respectively. We observe that the computation times of SPECTRODE largely depend on the aspect ratio, with increments of several orders of magnitude as $\gamma\to 1$, while they are much more uniform for MIXANDMIX and QuEST. As for the accuracy distributions, they are generally satisfactory for all methods, but MIXANDMIX seems to provide improved results near the lower edge for the MP case and throughout the support for the $\delta\delta$ case. The grid sizes used by each method for $\gamma=0.5$ have been $102/104$ by QuEST ($100$ plus some additional points to localize the support limits), $2930/5734$ by SPECTRODE and $1200/1200$ by MIXANDMIX, for the MP / $\delta\delta$ problems.

\begin{figure}[!htb]
\begin{minipage}{.33\textwidth}
\includegraphics[width=\textwidth]{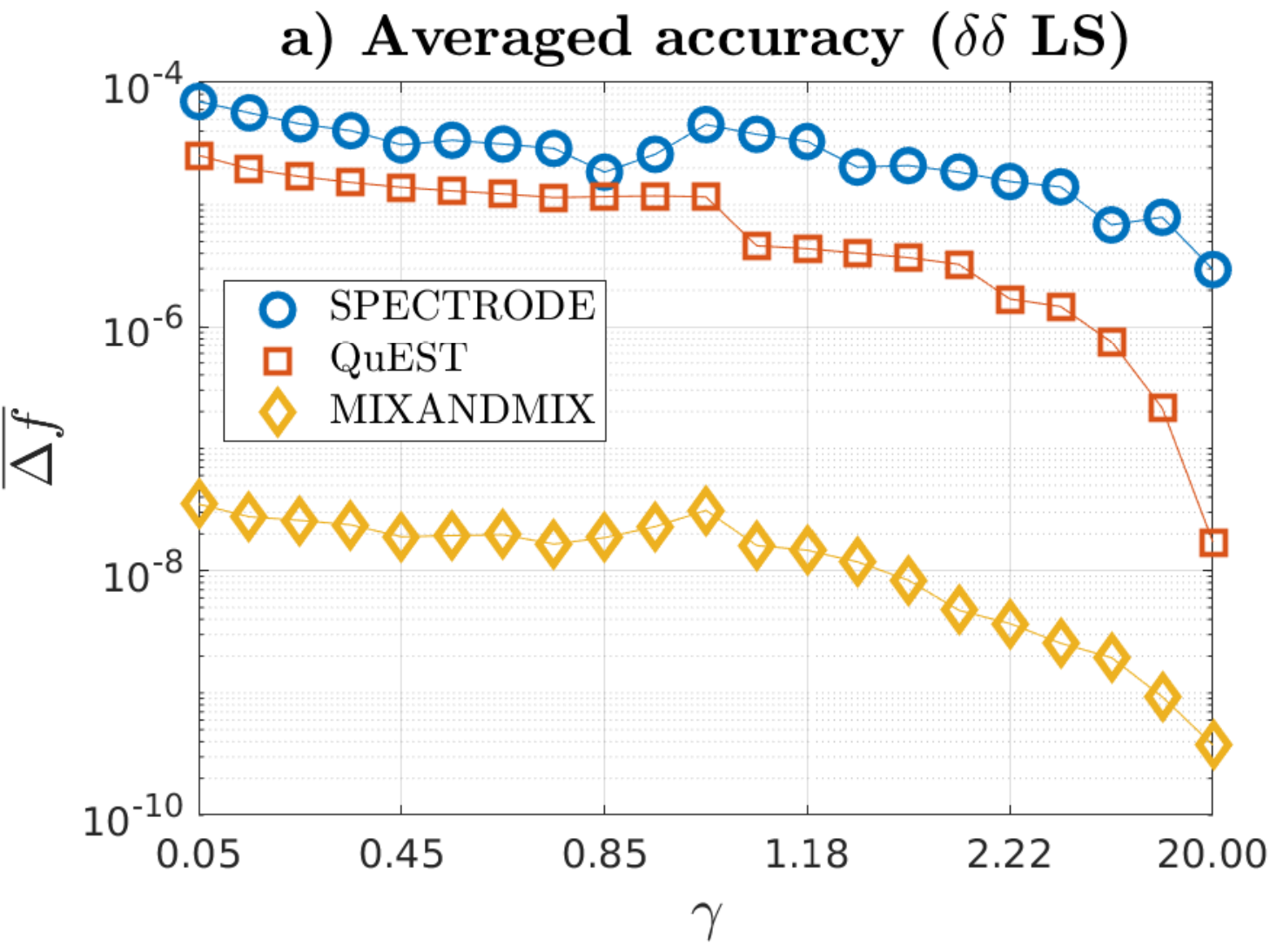}
\end{minipage}
\begin{minipage}{.33\textwidth}
\includegraphics[width=\textwidth]{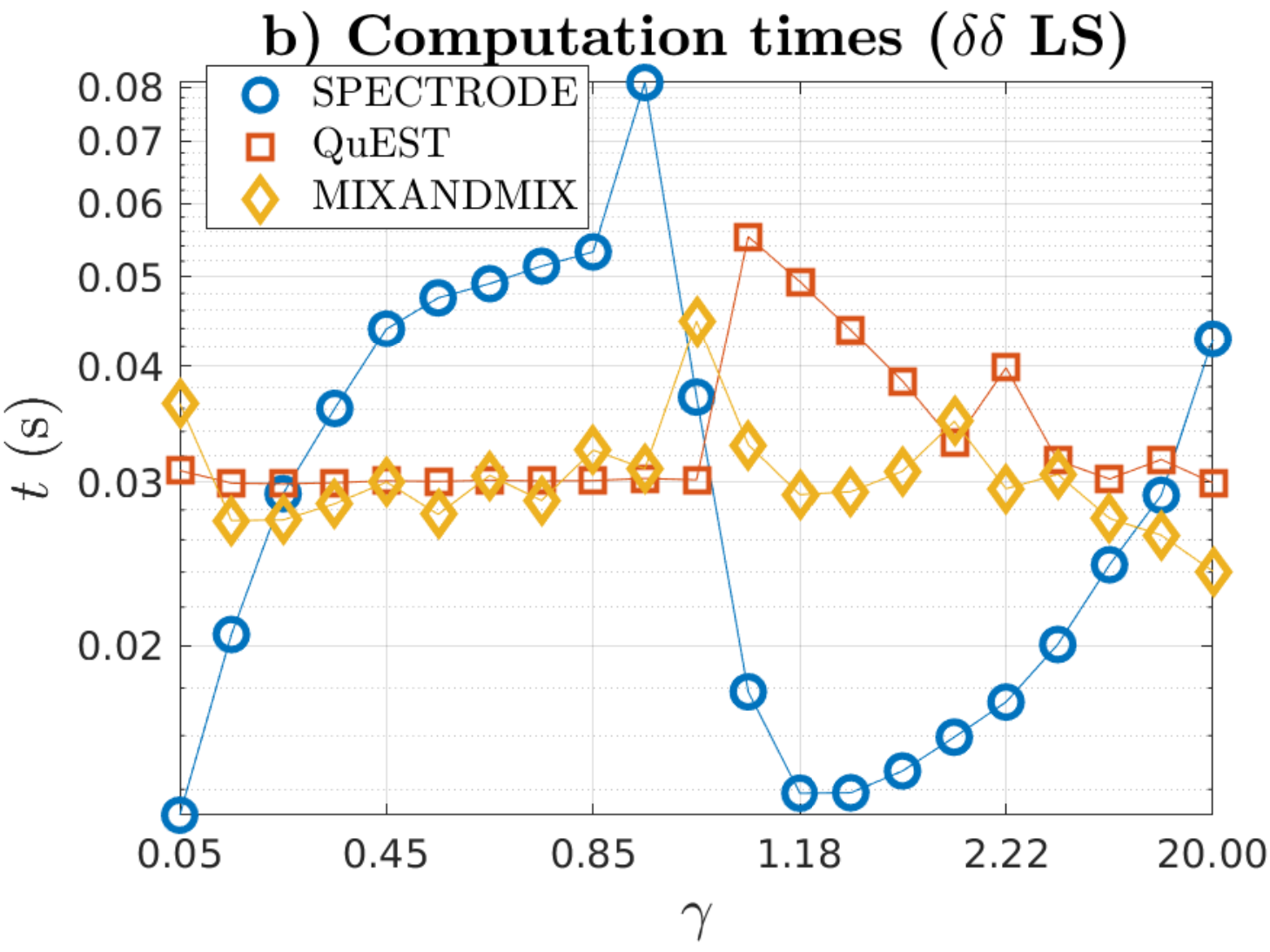}
\end{minipage}
\begin{minipage}{.33\textwidth}
\includegraphics[width=\textwidth]{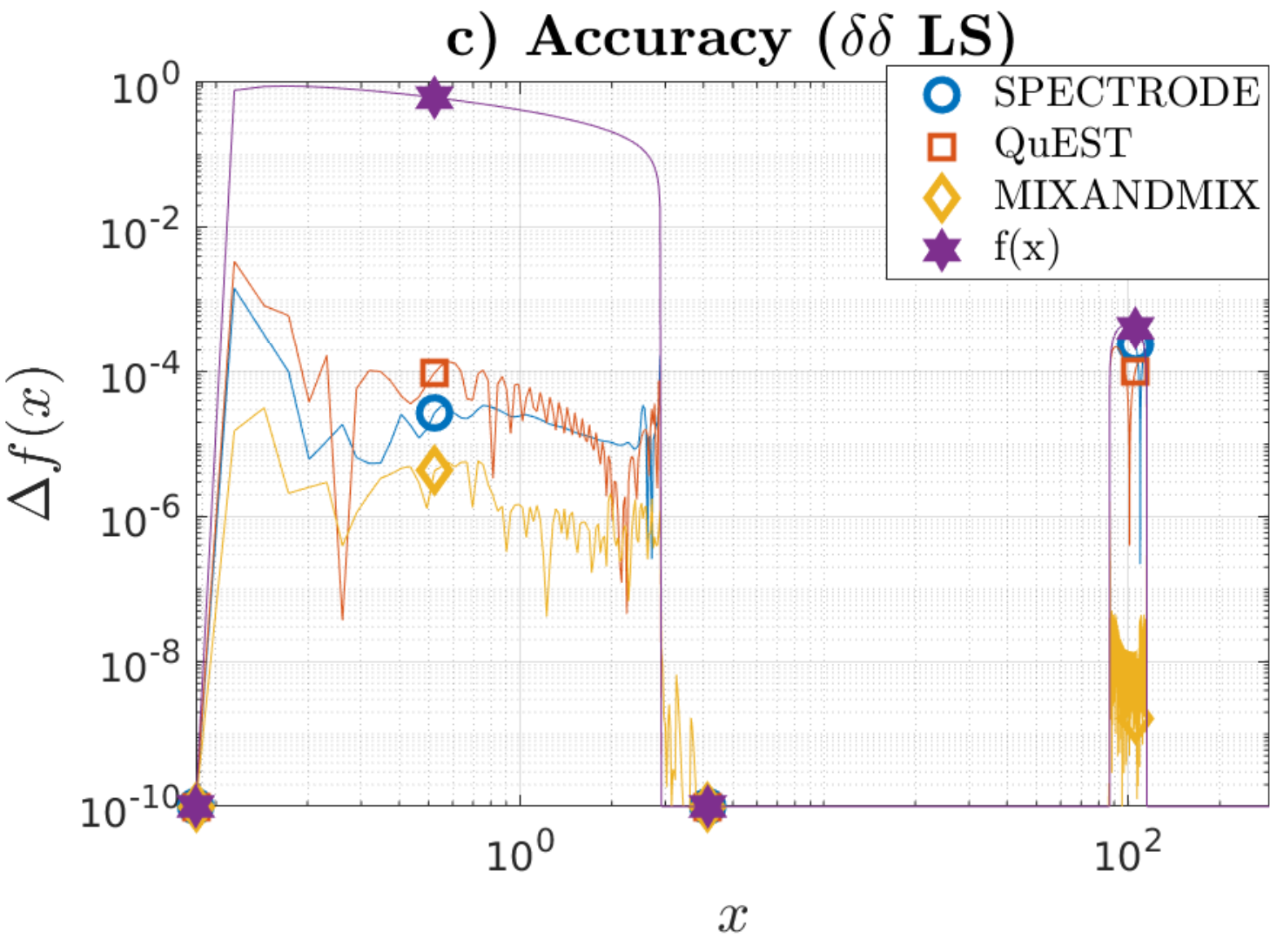}
\end{minipage}
\begin{minipage}{.33\textwidth}
\includegraphics[width=\textwidth]{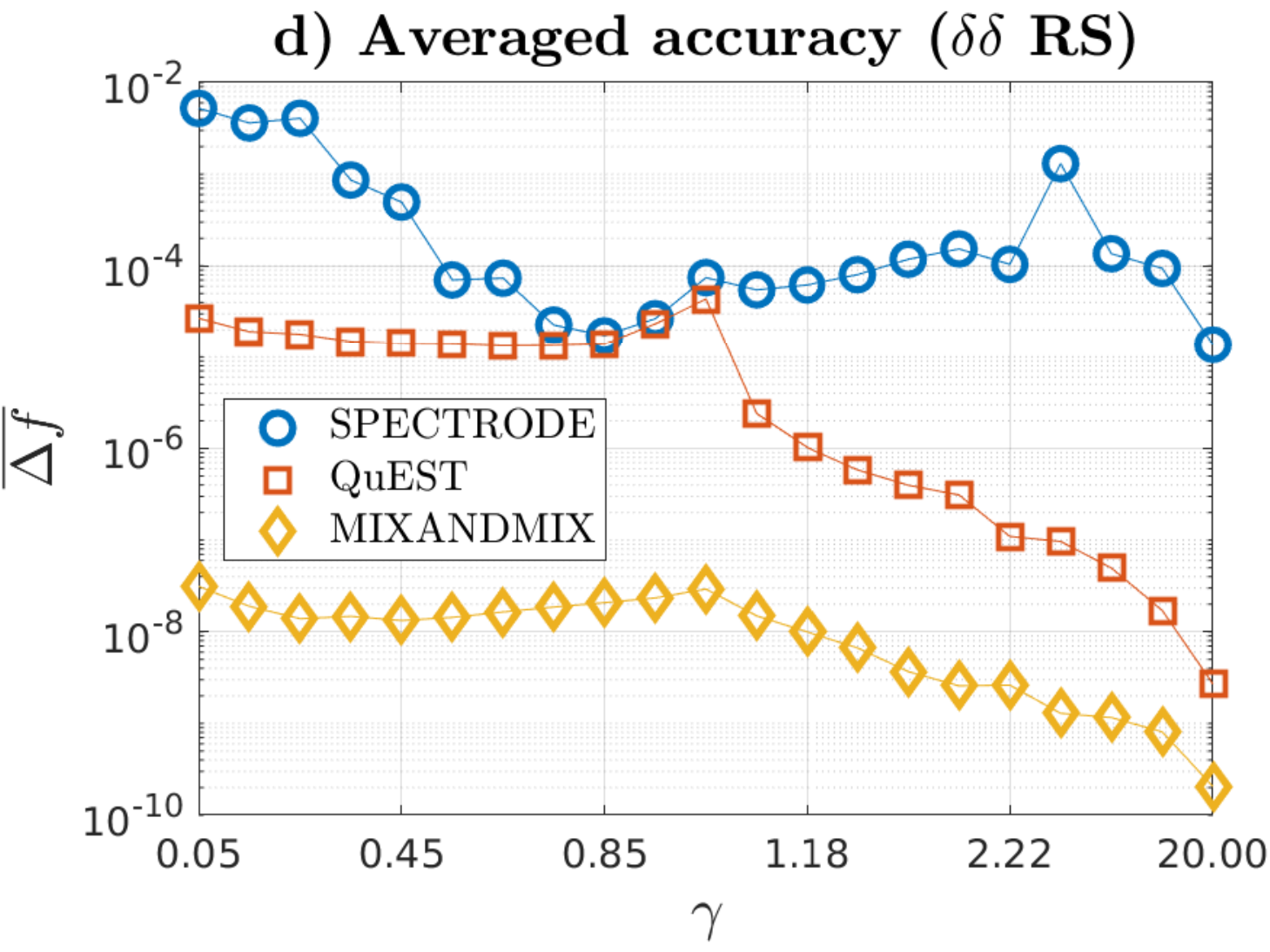}
\end{minipage}
\begin{minipage}{.33\textwidth}
\includegraphics[width=\textwidth]{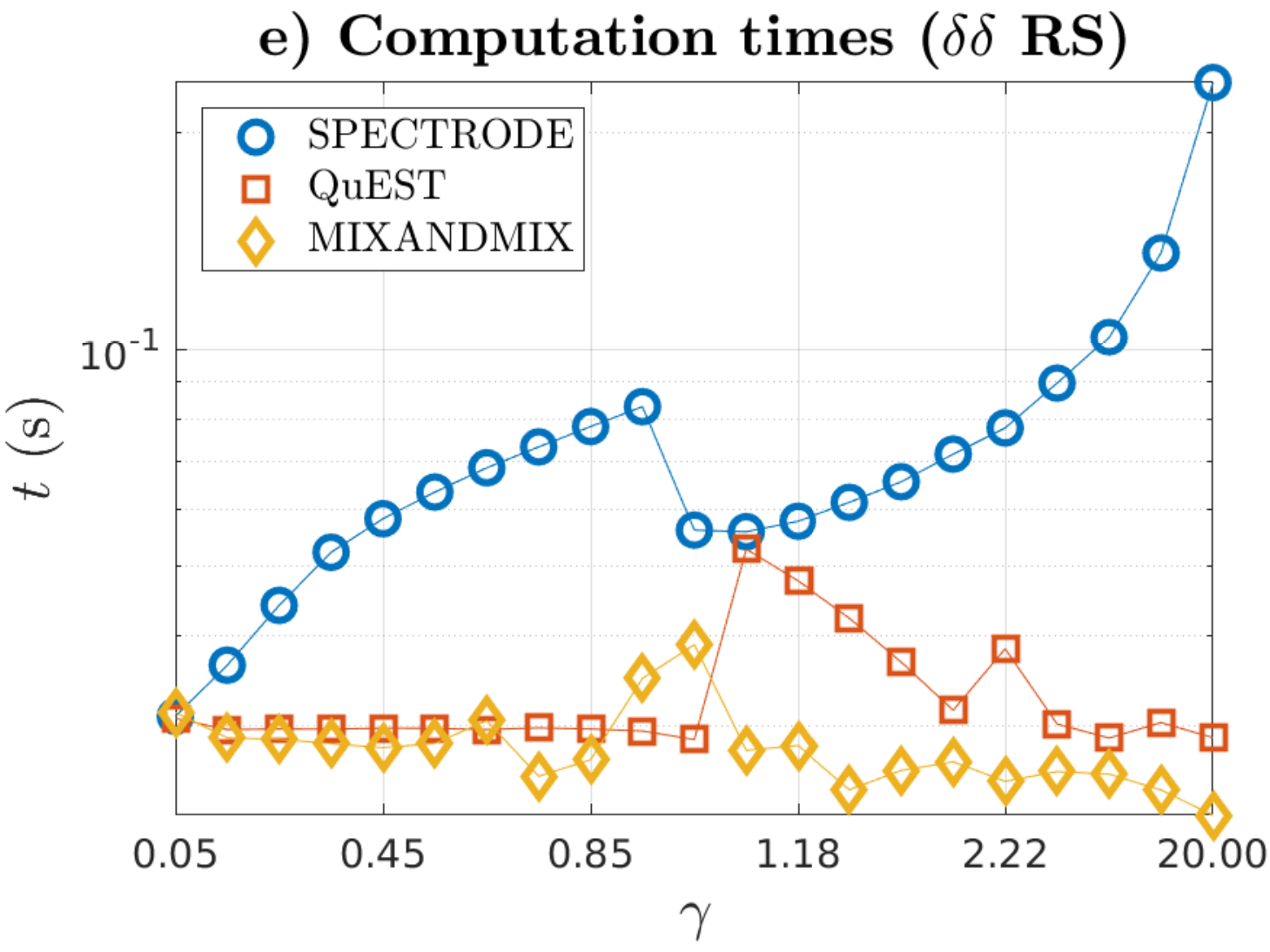}
\end{minipage}
\begin{minipage}{.33\textwidth}
\includegraphics[width=\textwidth]{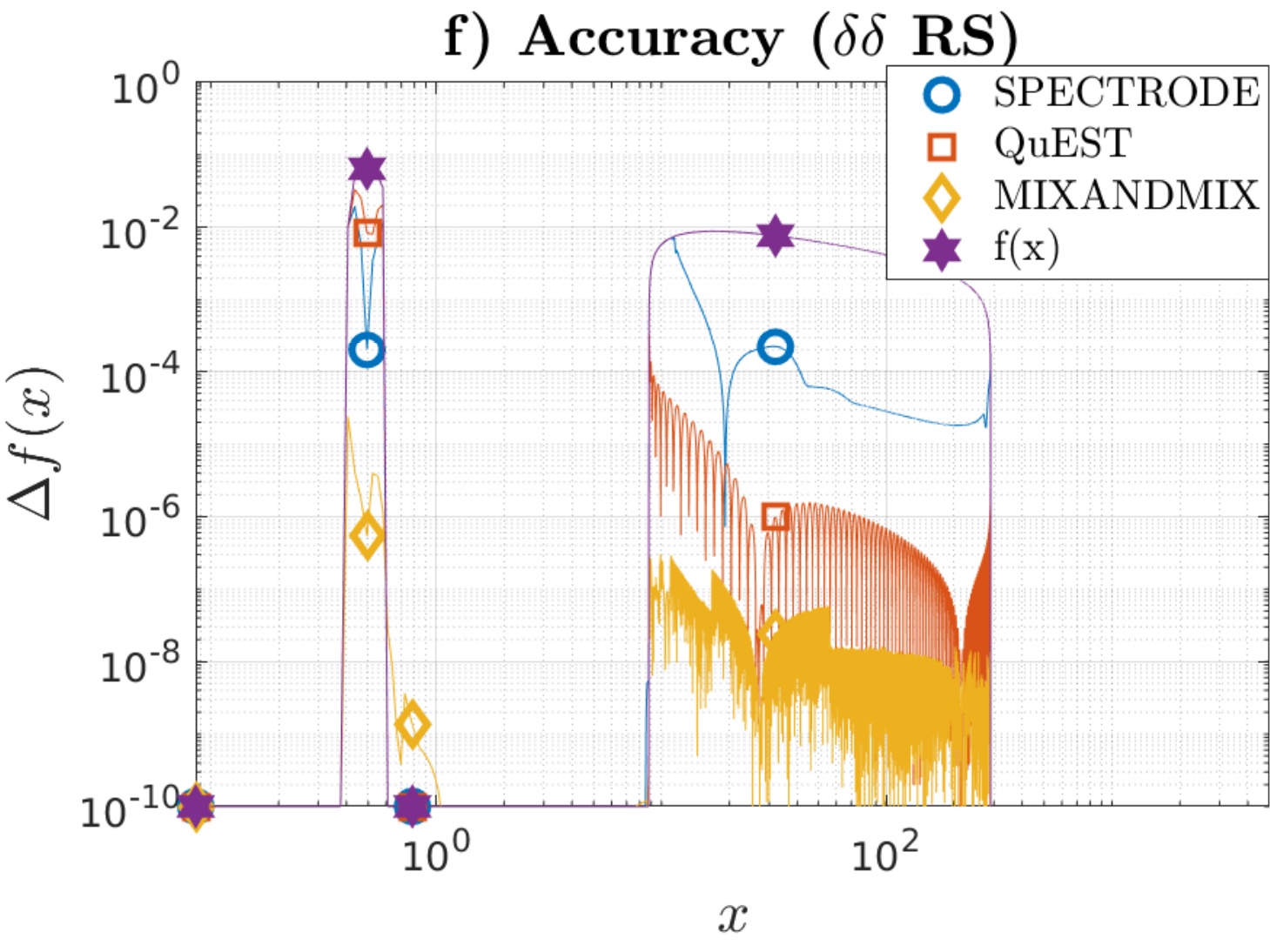}
\end{minipage}
\caption{\textbf{a,d)} Averaged accuracy, \textbf{b,e)} computation times, and \textbf{c,f)} accuracy throughout the domain ($\gamma=0.5$) for the $\delta\delta$ problem with $\boldsymbol{\lambda}=\{1,100\}$ skewed towards \textbf{a-c)} the smallest ($\mathbf{w}=\{0.99,0.01\}$) and \textbf{d-f)} largest ($\mathbf{w}=\{0.01,0.99\}$) eigenvalue.}
\label{fig:FIG4}
\end{figure}

MIXANDMIX is potentially limited by risks of failures at detecting all the segments comprising the distribution support. Thereby, we have conducted some experiments to test the support detection reliability in challenging scenarios. In Fig.~\ref{fig:FIG4} we cover analogous experiments to those in Fig.~\ref{fig:FIG3} for skewed $\delta\delta$ problems with $\boldsymbol{\lambda}=\{1,100\}$ with respective multiplicity ratios $\mathbf{w}=\{0.99,0.01\}$ (Figs.~\ref{fig:FIG4}a-c), left skewed (LS) problem, and $\mathbf{w}=\{0.01,0.99\}$ (Figs.~\ref{fig:FIG4}d-f), right skewed (RS) problem. Note that due to the scale differences in the population eigenvalue locations, results on the accuracy throughout the domain are more conveniently represented in a logarithmic scale. SPECTRODE parameters for approximately matched average computation times have now been modified to $\epsilon=10^{-4}/\epsilon=10^{-3}$ (LS, $\gamma<1/\gamma\geq 1$) and $\epsilon=10^{-3}/\epsilon=10^{-2}$ (RS, $\gamma<1/\gamma\geq 1$) while MIXANDMIX parameters could be kept the same as in previous experiments while still matching QuEST computation times. Once again, results in Figs.~\ref{fig:FIG4}a,d show an improvement of averaged accuracy by several orders of magnitude when using the MIXANDMIX method. Results on Figs.~\ref{fig:FIG4}c,f show that MIXANDMIX has been capable to approximate the support for both problems and to provide more accurate results than the other two methods almost everywhere within the support. In this case the grid sizes have been $104/104$ for QuEST, $6963/9130$ for SPECTRODE and $1248/1248$ for MIXANDMIX for the LS / RS problems.

\begin{figure}[!htb]
\begin{minipage}{.33\textwidth}
\includegraphics[width=\textwidth]{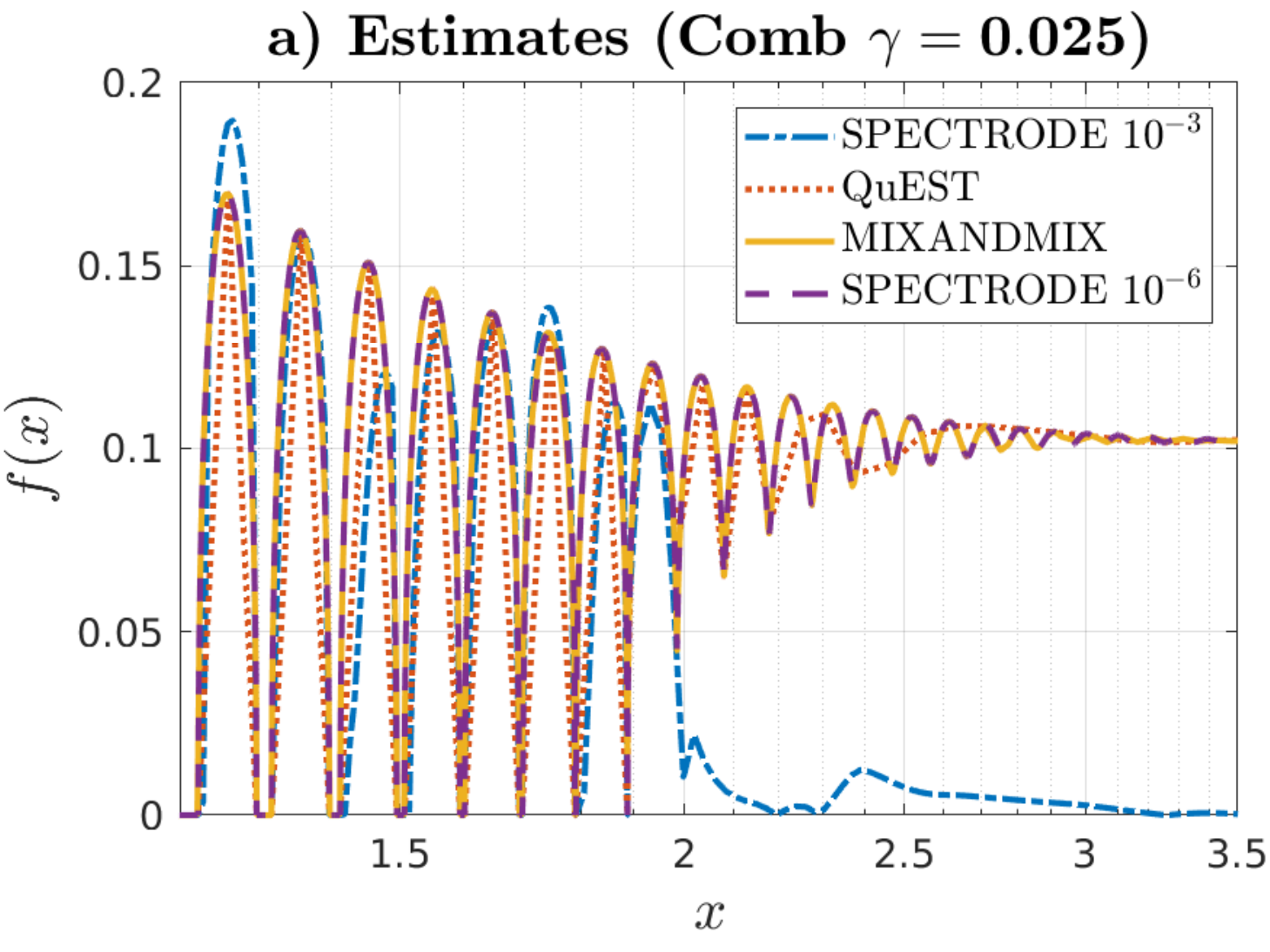}
\end{minipage}
\begin{minipage}{.33\textwidth}
\includegraphics[width=\textwidth]{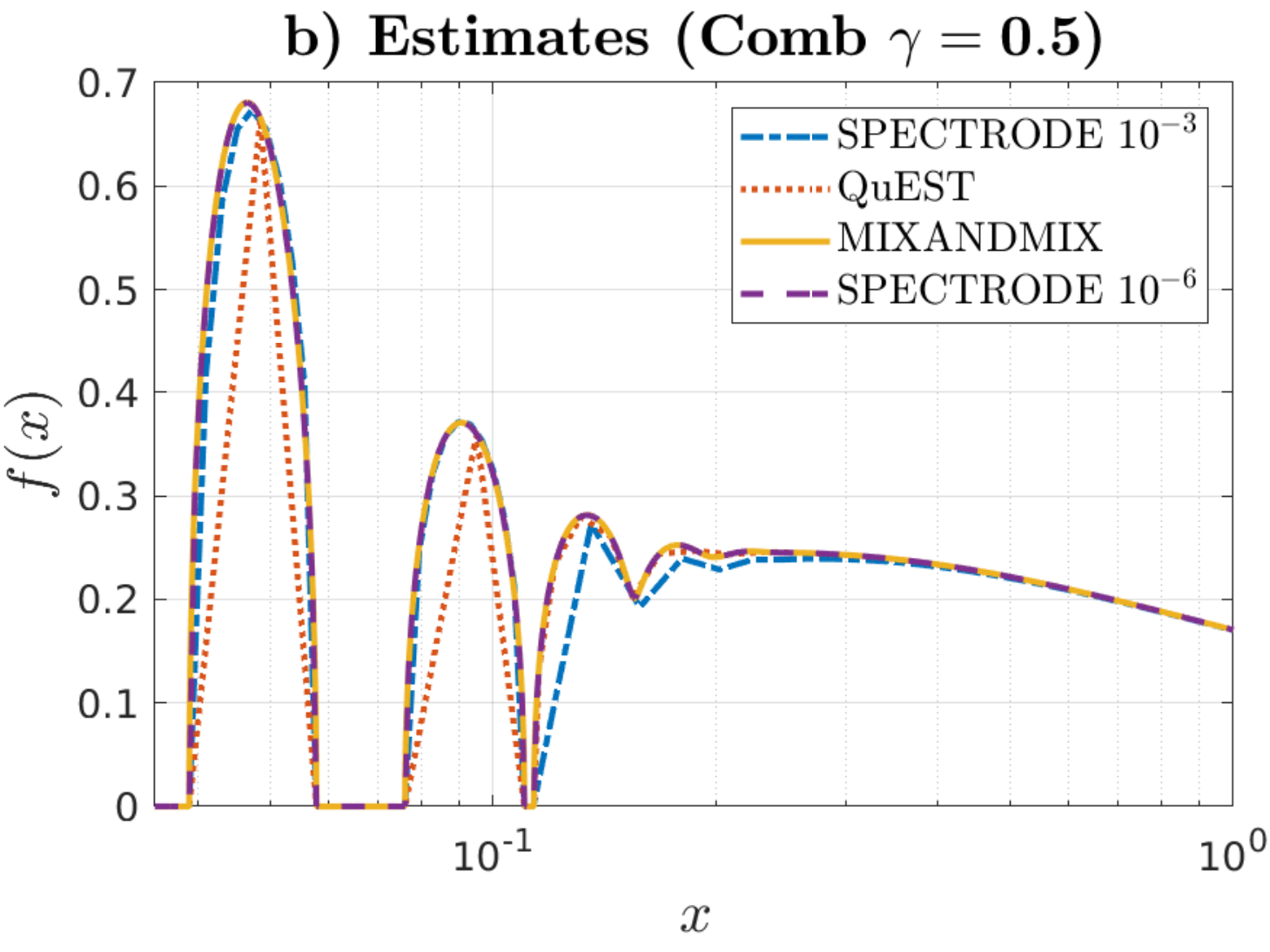}
\end{minipage}
\begin{minipage}{.33\textwidth}
\includegraphics[width=\textwidth]{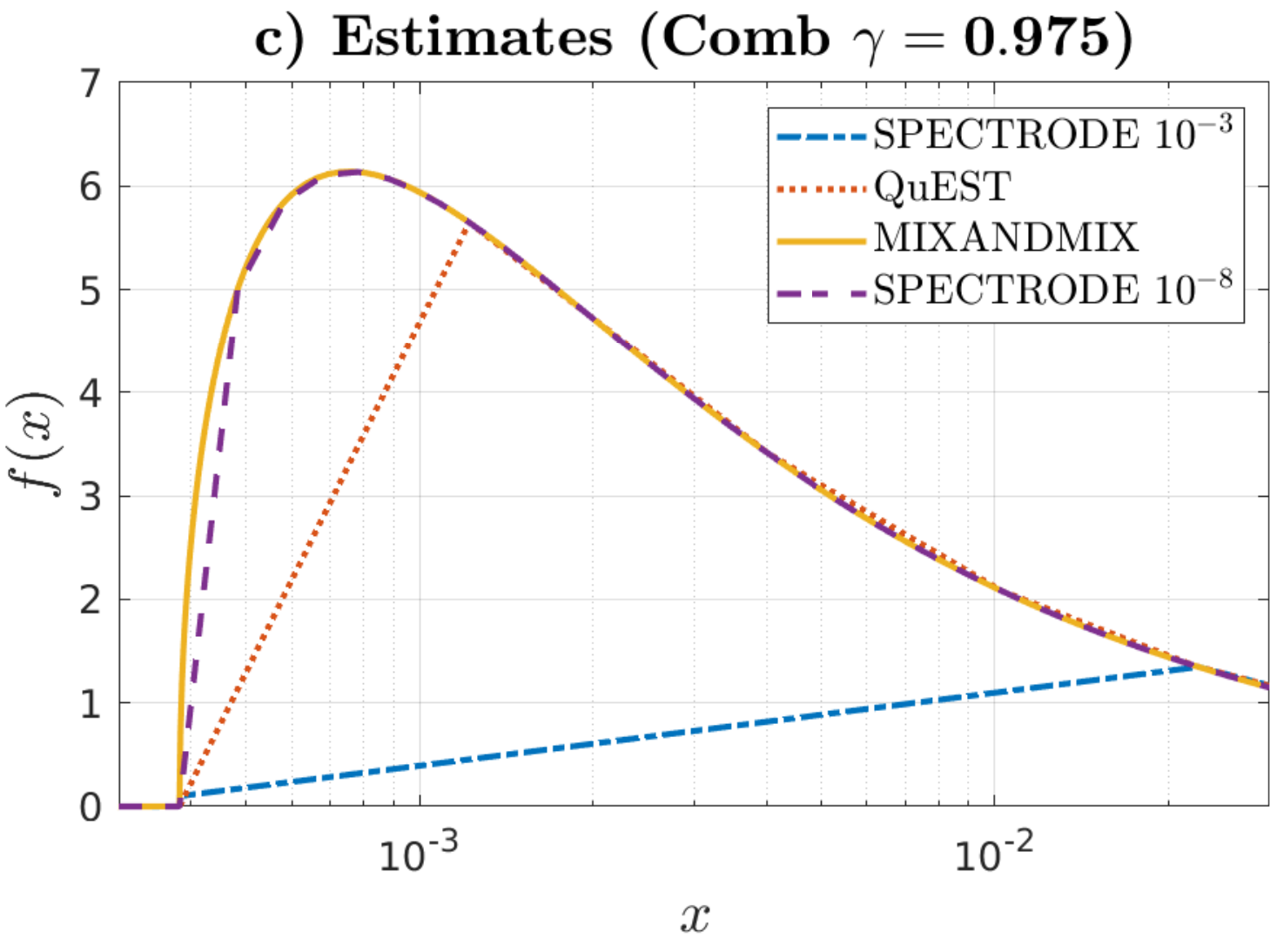}
\end{minipage}
\caption{ESD for the Comb problem with $100$ equiprobable point masses evenly distributed in the interval $[0.1,10]$. \textbf{a)} $\gamma=0.025$, \textbf{b)} $\gamma=0.5$ and \textbf{c)} $\gamma=0.975$.}
\label{fig:FIG5}
\end{figure}

In Fig.~\ref{fig:FIG5} we apply the three methods to a challenging comb-like~\citep{Dobriban15} problem. In this case the population eigenvalues come from $100$ equiprobable point masses uniformly distributed in the interval $[0.1,10]$, thus with a large number of components that differ by as much as two orders of magnitude. We show the results at selected areas of the support for $\gamma=0.025$, $\gamma=0.5$ and $\gamma=0.975$ respectively in Figs.~\ref{fig:FIG5}a-c, again using logarithmic scaling for improved visualization. When tuning the SPECTRODE parameter for similar computation times to those of QuEST and MIXANDMIX, which turned out to happen at $\epsilon=10^{-3}$, the visual impression is of limited performance. In contrast, MIXANDMIX appears to be powerful, showing strengths not only in detecting all the support intervals, but also in picking-up the density oscillations observed for $\gamma=0.025$ and the steeped lower edge for $\gamma=0.975$. We have quantitatively assessed this perception by running the final experiment including also the SPECTRODE results at the accuracy level where we could not visually detect any differences with MIXANDMIX results, $\epsilon=10^{-6}$ for both $\gamma=0.025$ and $\gamma=0.5$, or the maximum code accuracy, $\epsilon=10^{-8}$, for $\gamma=0.975$. Grid sizes of the different methods were $1950/1113/1216$ for SPECTRODE with matched computation times, $140/106/102$ for QuEST, $1248/1200/1200$ for MIXANDMIX, and $11678/19156/244296$ for SPECTRODE with improved accuracy. Results show no visual differences between MIXANDMIX and SPECTRODE with improved accuracy for $\gamma=0.025$ and $\gamma=0.5$, and more plausible functional shape of the former around the left edge for $\gamma=0.975$, even though approximately $200\times$ less computational resources were used. Finally, QuEST results show a remarkable ability to correctly determine the support intervals, but limitations to accurately approximate the spiked density areas or capture the density oscillations.

\subsection{ESD of population mixtures}

\label{sec:MIPO}

\begin{figure}[!htb]
\begin{minipage}{.33\textwidth}
\includegraphics[width=\textwidth]{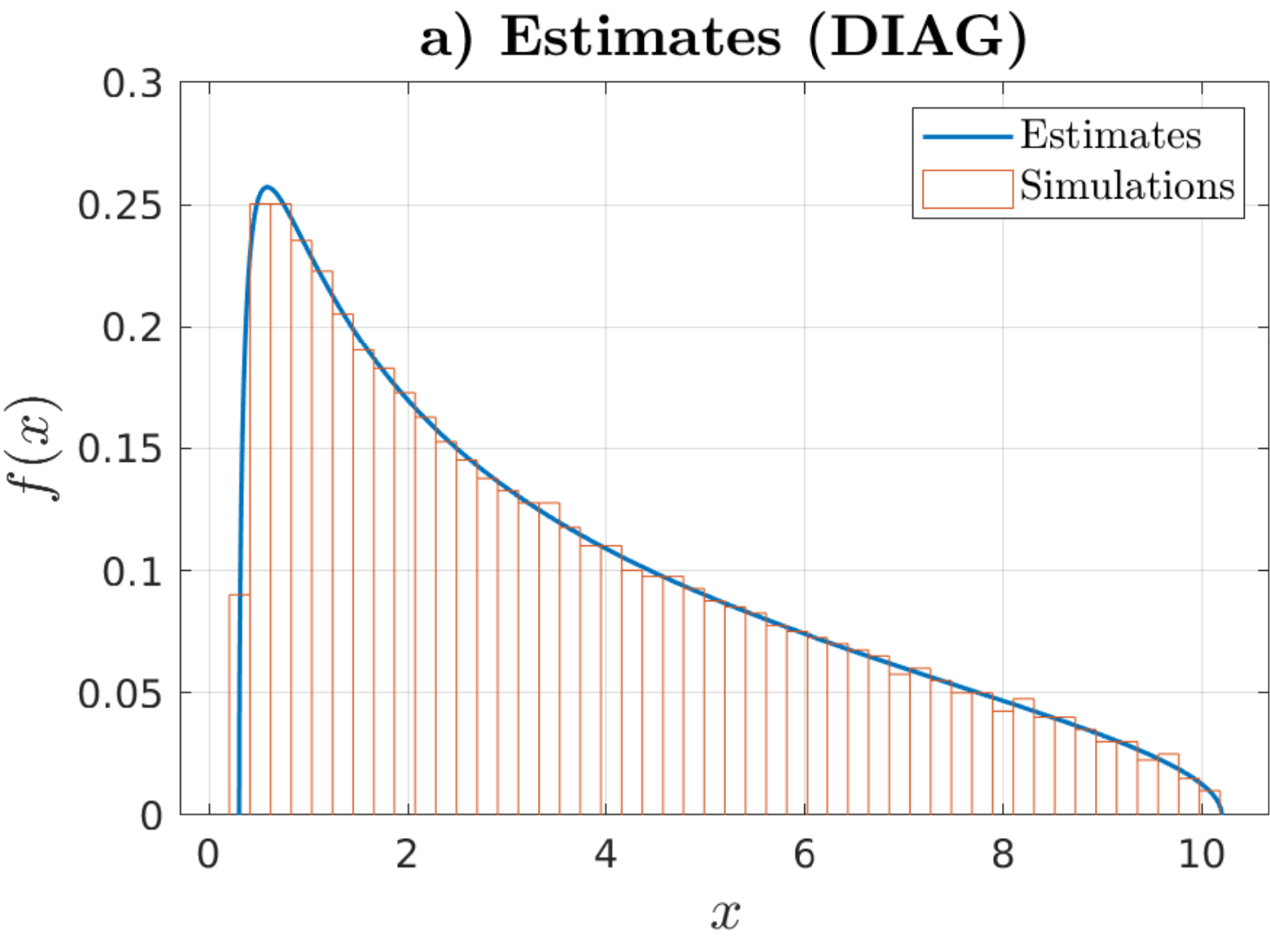}
\end{minipage}
\begin{minipage}{.33\textwidth}
\includegraphics[width=\textwidth]{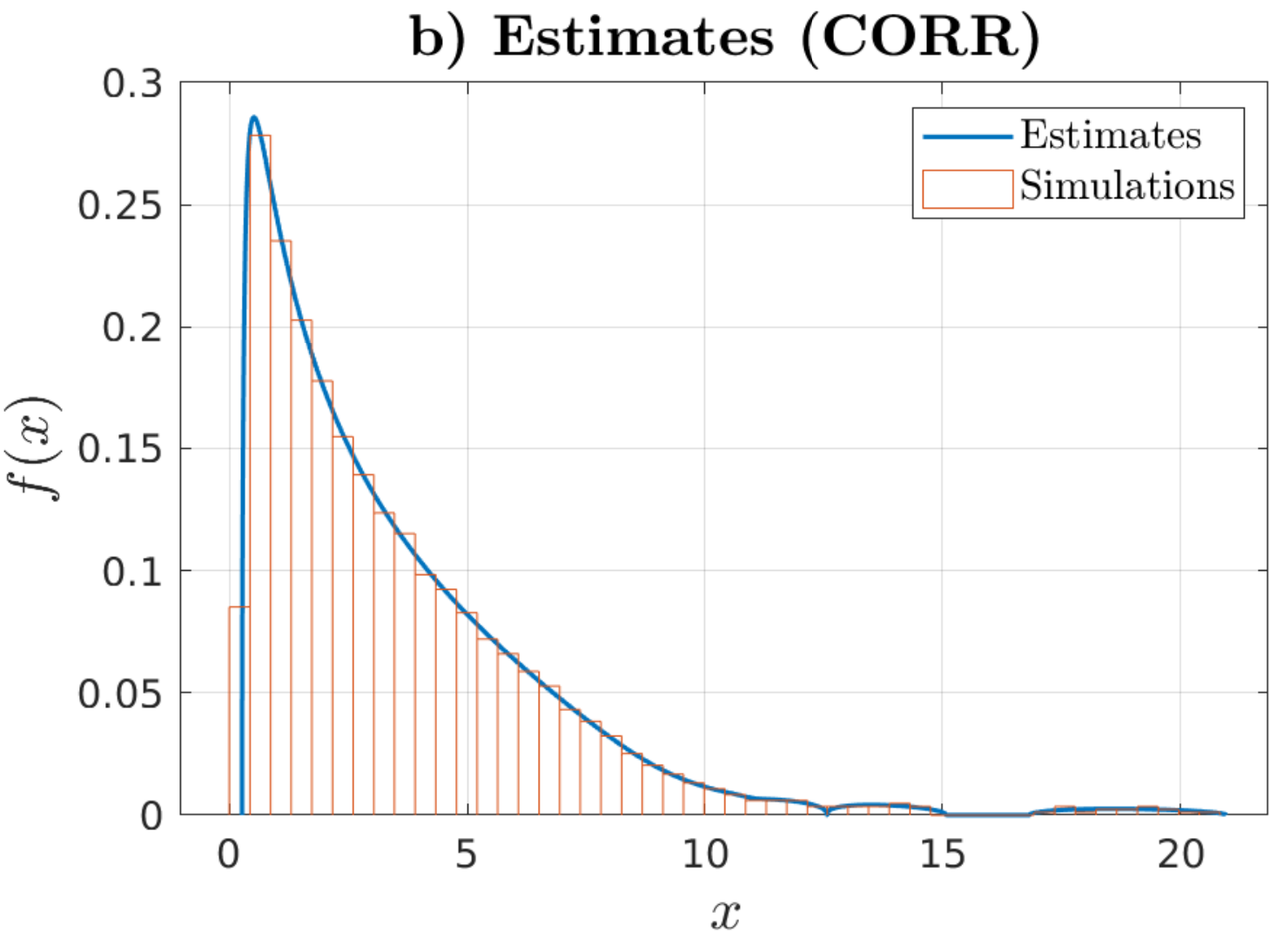}
\end{minipage}
\begin{minipage}{.33\textwidth}
\includegraphics[width=\textwidth]{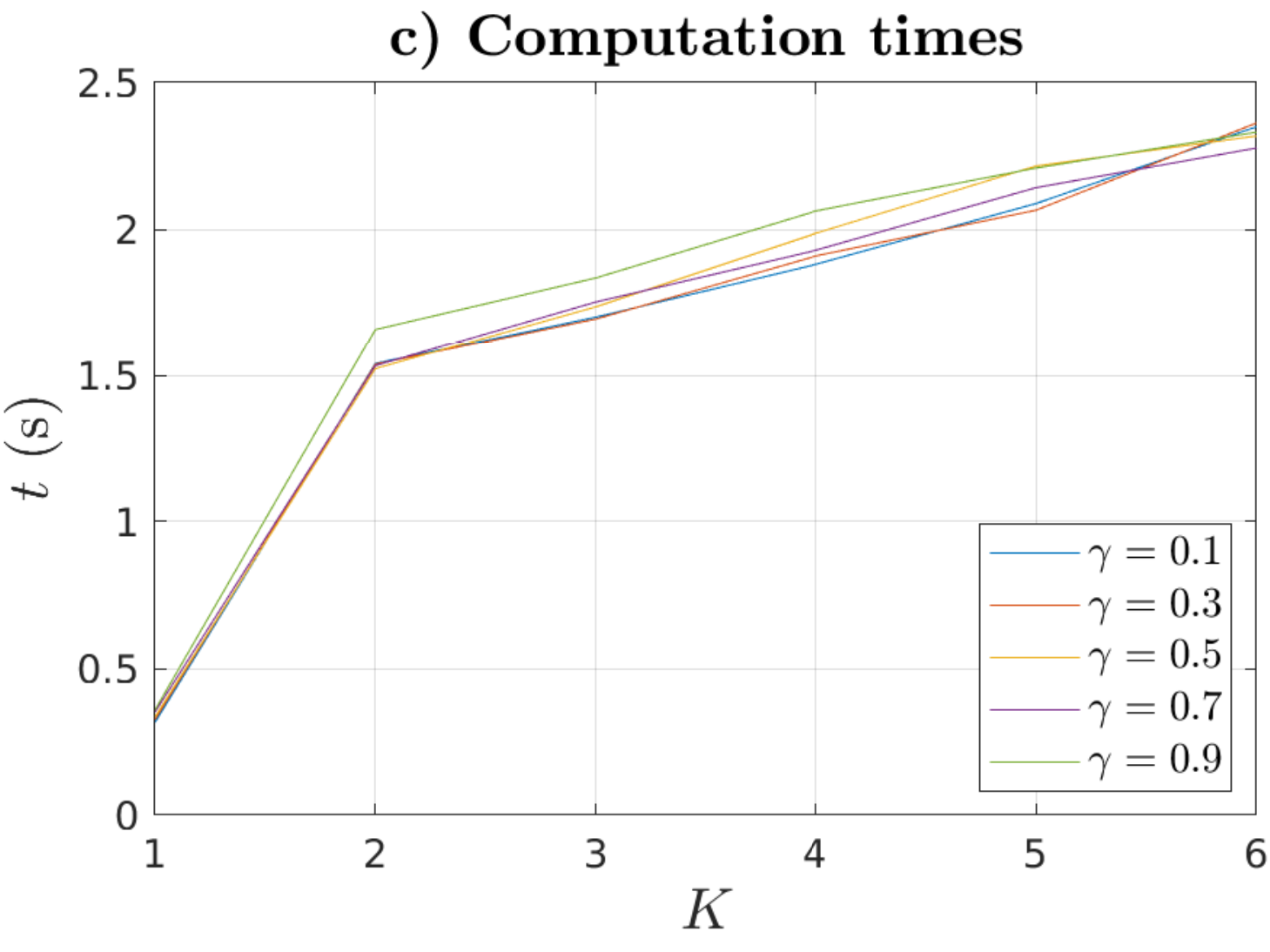}
\end{minipage}
\caption{Calculated ESDs and simulations for the \textbf{a)} DIAG and \textbf{b)} CORR problems. \textbf{c)} computation times for solving the CORR problem for different values of $\gamma$ and $K$.}
\label{fig:FIG6}
\end{figure}

In this Section we provide an illustration of the MIXANDMIX results for the mixture of populations model and the benefits of the GPU architecture. Fig.~\ref{fig:FIG6}a shows the calculated ESD for a mixture of populations with $K=6$ equiprobable populations, so $\alpha_k=1/K$, drawn from population covariances $\boldsymbol{\Lambda}_{M\text{(DIAG)}}^{k}=\Lambda_{mn\text{(DIAG)}}^k=((m+k)\bmod K)+1)\delta[m,n]$, i.e., a diagonal matrix whose elements in the diagonal grow from $1$ to $K$ with period $K$ with this pattern being shifted across the populations. Fig.~\ref{fig:FIG6}b extends this DIAG problem to non-diagonal population covariances using $\boldsymbol{\Lambda}_{M\text{(CORR)}}^{k}=\Lambda_{mn\text{(CORR)}}^k=\rho^{|m-n|^l}\sqrt{\Lambda_{mm\text{(DIAG)}}^k\Lambda_{nn\text{(DIAG)}}^k}$, $\rho<1$, $l>0$, so this CORR problem reduces to the DIAG problem when $l\to\infty$. Namely, Fig.~\ref{fig:FIG6}b  shows the results for $\rho=0.2$, $l=0.25$ and $\gamma=0.5$, here with $M=120$ for convenience. We can see that calculations are in agreement with simulations (with these being obtained using the biggest matrix sizes that we would fit in our GPU memory) for both the DIAG and CORR problems, with the CORR results showing a larger spectral dispersion due to the larger condition number of the non-diagonal matrix. Finally, Fig.~\ref{fig:FIG6}c shows the GPU computation times for the CORR problem for a number of populations ranging from $K=1$ to $K=6$. First, we can appreciate a significant penalty when moving from $K=1$ to $K=2$, as that switches the problem from solving a single equation based only on the eigenvalues to a system of equations based on the whole structure of the covariance matrices. Second, we observe that for $K\geq 2$ the computation times grow with $K$ following a lower than $1$ ratio. This is to be attributed to an increased degree of parallelization of the GPU implementation for bigger problems and a stable fixed point Anderson acceleration in the multidimensional case.

\section{Discussion}

\label{sec:DISC}

We have presented a set of numerical techniques to aid in the computation of the ESD in a mixture of populations model. These include the use of Anderson mixing to accelerate the fixed point iterations, homotopy continuation for robust convergence to the right optimum, adaptive grid construction to efficiently detect the support and approximate the distribution, and a parallel architecture to tackle the increased computational demands in this setting. Results have shown that our method offers favorable practical efficiency-accuracy tradeoffs when compared with related approaches while being able to address more general models than available tools. 

Our tool has focused on the model described in~\cite{Wagner12} but it could straightforwardly be adapted and optimized to address several analogous models in the literature, such as those mentioned in~\S~\ref{sec:INTR}. Nevertheless, we have not contemplated even more general models such as the Kronecker model~\citep{Zhang13}, with a recent contribution for the separable case in~\cite{Leeb19}, or more general couplings between the matrix elements~\citep{Wen11,Lu16}. However, these models generally involve the solution of more intricate systems of nonlinear equations with more auxiliary functions, but with strong functional resemblances to the system we have studied here, so there is potential to reuse or adapt our tools to tackle them. The population model discussed in this paper admits a free probability~\citep{Mingo17} based formulation given by
\begin{equation}
\label{ec:FPFP}
f_\gamma^{\prescript{}{K}{\boldsymbol{\Lambda}}^{}_{M}}=\displaystyle\op_{k=1}^K\alpha_k(f_0^{\boldsymbol{\Lambda}_M^k}\boxtimes f^{\mathbf{I}}_{\gamma /\alpha_k}),
\end{equation}
where $\boxtimes$ represents the free multiplicative convolution and $\op$ the free additive convolution over the summands. This implies that the machinery described in~\cite{Belinschi17} could be used to obtain the system of equations in~\S~\ref{sec:INCO}. However, the models in~\cite{Belinschi17} are more general, including self-adjoint polynomials not necessarily built from a combination of Mar\v{c}enko-Pastur and atomic distributions, but simply from asymptotically freely independent matrices. Thus, investigations are required to discern under what conditions our numerical tools can be extended to cope with these models.

The proposed technique does not fully exploit the existing descriptions used for support detection in the single population model because we are not aware of any such descriptions for mixtures of populations (although recent results such as those in~\cite{Bao19,Ji19} may pave the way for them). Anyhow, the gridding procedure has shown a robust behaviour in challenging practical scenarios, even when compared with methods that exploit the properties of the support. The main reason is the introduction of a logarithmic grid, that enables efficient support searches at different spectral scales. This has been synergistically combined with an adaptive grid refinement making use of the second order derivative of the density (with a bias towards the upper edge information) generally with more efficient approximations than provided by previous methods. Although this grid refinement criterion has proven effective for all the tested cases, generally providing finer spectral resolvability than previously proposed methods, other criteria may be more appropriate for different applications. In this regard, we should mention that our software is built on top of a generic grid refinement method that allows to test other possibilities by simply defining different criteria for grid cell subdivision, with some exemplary alternatives already included in the code. In addition, support detection correctness can generally be inferred from the numerical integration of the estimated measures, which is provided as an output.

Another difference with previous approaches is the dependence of the method on more parameters. Although this may add an extra degree of complexity for users, we have observed good behaviour for all test cases without resorting to parameter tuning, so the tool should be readily usable for many applications. From a different perspective, the combined inspection of simulations and manipulation of these parameters may allow to fine tune the method in most challenging scenarios, some of them, as shown in some of the experiments in~\S~\ref{sec:COLI}, not being adequately covered by the reduced flexibility of related approaches. In summary, grid detection robustness can be improved by increasing $M^{\text{(i)}}$, robustness of approximation by increasing $\xi^0$ and/or decreasing $\beta$ and accuracy by decreasing $\epsilon$ and/or increasing $L$. However, a line for future research could involve the incorporation of refined techniques for parameter selection. A possibility could be to investigate more efficient and robust interplays between homotopy continuation and nonlinear acceleration, for instance based on adaptive regularization schemes for nonlinear acceleration~\citep{Scieur19}.

\section{Conclusions}

\label{sec:CONC}

This work has introduced a set of techniques to compute the ESD in a mixture of populations model. A generic procedure using only the functional form of the fixed point equations relating the population and limiting empirical distributions has been proposed. Efficient convergence is achieved by Anderson acceleration and homotopy continuation and novel strategies for grid construction have been provided. The method has compared well with related proposals in the literature which, to our knowledge, are only capable to address more restricted models. By providing this detailed description of our solution, we expect our distributed tool to be of practical interest for statisticians working in the field.

\section*{Acknowledgments}

This work received funding from the European Research Council under the European Union's Seventh Framework Programme (FP7/20072013/ERC grant agreement no. [319456] dHCP project). The research was supported by the Wellcome/EPSRC Centre for Medical Engineering at King's College London [WT 203148/Z/16/Z]; the Medical Research Council [MR/K006355/1]; and the National Institute for Health Research (NIHR) Biomedical Research Centre based at Guy's and St Thomas' NHS Foundation Trust and King's College London. The views expressed are those of the author and not necessarily those of the NHS, the NIHR or the Department of Health. The author also acknowledges the Department of Perinatal Imaging \& Health at King's College London and advice and support from Edgar Dobriban, Jo V. Hajnal and Daan Christiaens.

\clearpage

\bibliographystyle{apalike}
\bibliography{ESD}

\clearpage


\end{document}